\documentclass[12pt]{article}
\usepackage{latexsym,epsfig,graphicx,amsmath,amssymb,amscd,undertilde,multirow,chicago,psfrag,paralist,dsfont,algorithm,algpseudocode}
\usepackage{mathtools}
\usepackage{bm}
\usepackage{hyperref}
\usepackage[titletoc]{appendix}

\usepackage[american]{babel}

\usepackage{epstopdf}

\newcommand{\betavec}{{\boldsymbol{\beta}}}

\newcommand{\diag}{\mathop{\mathrm{diag}}}

\newcommand{\EE}{{\mathbf E}}

\newcommand{\etavec}{{\boldsymbol{\eta}}}

\newcommand{\gammavec}{{\boldsymbol{\gamma}}}

\newcommand{\muvec}{{\boldsymbol{\mu}}}

\newcommand{\N}{{\textrm{N}}}

\newcommand{\onevec}{{\boldsymbol{1}}}

\newcommand{\sigmavec}{{\boldsymbol{\sigma}}}

\newcommand{\thetavec}{{\boldsymbol{\theta}}}

\newcommand{\KL}{\textrm{KL}}

\newcommand{\xvec}{\boldsymbol{x}}

\newcommand{\zerovec}{{\boldsymbol{0}}}

\newcommand{\Dcal}{\mathcal{D}}
\newcommand{\Lcal}{\mathcal{L}}
\newcommand{\MVN}{{\rm MVN}}
\newcommand{\IGAM}{{\rm IGAM}}

\newcommand{\tran}{^\top}

\begin{document}
	\title{The Use of Variational Inference for Lifetime Data with Spatial Correlations}
	
		\author{Yueyao Wang$^{1,2}$, Yili Hong$^3$, Laura Freeman$^3$ and Xinwei Deng$^3$\footnote{Address for correspondence: Xinwei Deng, Professor, Department of Statistics,
						Virginia Tech, Blacksburg, VA 24061 (E-mail: xdeng@vt.edu).} \\
				\footnotesize $^1$School of Statistics and Mathematics, \\
				\footnotesize Zhejiang Gongshang University, Hangzhou, China 310018\\
				\footnotesize$^2$Collaborative Innovation Center of Statistical Data  Engineering Technology \& Application,\\
				\footnotesize Zhejiang Gongshang University,Hangzhou, China 310018\\
				\footnotesize $^3$Department of Statistics, Virginia Tech, Blacksburg, VA 24061
			}

	\date{}
	
	\maketitle
	
	\begin{abstract}
		Lifetime data with spatial correlations are often collected for analysis in modern engineering, clinical, and medical applications. For such spatial lifetime data, statistical models usually account for the spatial dependence through spatial random effects, such as the cumulative exposure model and the proportional hazards model. For these models, the Bayesian estimation is commonly used for model inference, but often encounters computational challenges when the number of spatial locations is large. The conventional Markov Chain Monte Carlo (MCMC) methods for sampling the posterior can be time-consuming. In this case-study paper, we investigate the capability of variational inference (VI) for the model inference on spatial lifetime data, aiming for a good balance between the estimation accuracy and computational efficiency. Specifically, the VI methods with different divergence metrics are investigated for the spatial lifetime models. In the case study, the Titan GPU lifetime data and the pine tree lifetime data are used to examine the VI methods in terms of their computational advantage and estimation accuracy.

		\textbf{Key Words}: $\alpha$-divergence, Accelerated Failure Time Model, Bayesian Inference, Hamiltonian Monte Carlo, Proportional Hazards Model, Spatial Random Effects.
	\end{abstract}
	
	\newpage
	\section{Problem Description}\label{sec:vi.introuduction}
	
	Spatially correlated lifetime data are commonly observed in engineering, medical, and biostatistics fields, when the data are collected over spatial locations, such as clinical sites and geographical regions. For example, \shortciteN{ostrouchov2020gpu} provide a detailed analysis of Cray XK7 Titan supercomputer graphics processing unit (GPU) lifetime data. Inside the Titan supercomputer, GPU nodes are regularly arranged in the server room and present a spatial pattern. For such data, there exists unobserved heterogeneity in observations among different locations. Besides, the lifetime data in adjacent locations may be correlated, because closer locations can share similar environmental characteristics. Thus, modeling the spatial correlation is crucial for understanding the time-to-event data across locations.
	
	Existing approaches incorporate the spatial correlation via extending lifetime models with spatially correlated random effects. The spatial random effects are assumed to follow a certain distribution, and the correlations between random effects are described by the variance-covariance matrix of the distribution. For example, \citeN{henderson2002modeling} investigate possible spatial variation in the survival of adult Leukemia patients based on Bayesian hierarchical proportional hazards (PH) and proportional odds models. The study of \citeN{banerjee2003frailty} employs the PH model with gamma frailties to explain the pattern of infant mortality among counties while accounting for important covariates and spatial correlations. \citeN{diva2008parametric} study the joint modeling of multiple cancers while accounting for spatial correlations. 
	\citeN{zhao2009mixtures} illustrate a flexible spatial frailty lifetime modeling with an analysis of spatially oriented breast cancer data. \shortciteN{pan2014bayesian} study the interval-censored spatial survival under a proportional hazards frailty model.  \shortciteN{li2015survival} present a semi-parametric PH model for the lifetime of pine trees across 182 sites in the United States. \shortciteN{jie2022gpu} use mixture distributions within the lifetime regression model to study the potential effect of spatial locations on the Titan GPU lifetime with competing risks.
	In the above studies, the Bayesian framework is often adopted for the model inference. 
	Because the number of spatial locations can be large with over thousands of locations, the posterior distribution is often intractable. Markov chain Monte Carlo (MCMC) approaches are used in many studies to generate samples from the posterior. Recently, Hamiltonian Monte Carlo (HMC), as an efficient MCMC method in dealing with high-dimensional distributions, receives great attention. Various algorithms, such as No-U-Turn Sampler (NUTS), are deployed for HMC \cite{hoffman2014no}.
	
	With a massive amount of data and complex models, the computational cost of HMC can, however, limit its applications. Thus, we are motivated to investigate an approximation inference approach, namely variational inference (VI), for spatial lifetime model inferences. VI is an optimization-based approach to approximate intractable distributions. VI approximates the complicated posterior with a simple variational distribution by minimizing the metrics that quantify the closeness between two distributions. Commonly used metrics include the Kullback-Leibler (KL) divergence, the $\alpha$-divergence, and the $\chi-$divergence \cite{wan2020f}. 	\citeN{viblei} provide a thorough review of VI. Minimizing the KL divergence between the posterior and a variational distribution that belongs to a mean-field family is a standard VI procedure in many studies. But the parameter independence assumption of the mean-field family limits the flexibility of the variational distribution, and the KL divergence can lead to the underestimation of the posterior variance. Structured VI that allows dependencies between parameters is one direction to introduce flexibility to variational distributions. For example, \citeN{hoffman2015structured} allow arbitrary dependencies between global and local hidden variables for better parameter estimations. \shortciteN{kucukelbir2017automatic} consider Gaussian variational families with non-diagonal covariance. \citeN{ranganath2016hierarchical} propose hierarchical Gaussian process priors to allow the correlation between parameters through a more flexible and structured approximation distribution. The other direction is to use different divergence metrics, such as the $\alpha$-divergence as in \shortciteN{hernandez2016black} to allow flexible behavior of the variational distribution.


	Some papers apply VI to lifetime data inference. \citeN{kim2018variational} apply VI to a Gaussian process survival analysis model, which is an extension of the Cox PH model. \shortciteN{bovskoski2021variational} use variational Bayes to estimate parameters in an unemployment problem. \citeN{jung2021bayesian} propose a population-level inference in electronic health records with stochastic VI. \citeN{xiu2020variational} study the variational inference capability of model fitting with individual survival analysis. However, these works did not consider the VI for lifetime data with spatial correlations. Thus, it is worth investigating the capability and computational efficiency of VI in spatially correlated lifetime data and conducting the performance comparison between HMC and variational inference.

	In this paper, motivated by the importance of computational efficiency, we focus on investigating the performance of VI for spatial lifetime models regarding computing time and inference capability. The rest of this paper is organized as follows. Section \ref{sec:data_prep} provides a detailed description of the Titan GPU dataset and the pine tree lifetime dataset. Section \ref{sec:methods} introduces the models used for spatially correlated lifetime data as well as a general VI inference procedure.  The numerical results are presented in Section~\ref{sec:casestudy}. Finally, Section~\ref{sec:vi.con} summarizes the contribution and conclusion of this work.
	\section{Data Collection and Processing} \label{sec:data_prep}

	\subsection{Titan GPU Lifetime Data}\label{sec:vi.gpu.description}
	
	The original Titan GPU lifetime dataset is publicly available in \shortciteN{ostrouchov2020gpu}, where the serial numbers, locations, failure type, and life spans are extracted from inventory runs and failure event log records from 2012-2019. During the service time of the Titan supercomputer, a GPU node can fail multiple times due to different reasons. In this study, since our goal is to investigate the computational efficiency of VI, we only take into account the first error time of each GPU node. The modified version of the dataset as described in \shortciteN{jie2022gpu} is adopted. In particular,  one main failure type, off the bus (OTB), is focused on. GPU nodes that failed due to OTB are recorded as failed, and other failure modes, including censored nodes, are recorded as censored. 
	
\begin{figure}
		\centering
		\includegraphics[width = \linewidth]{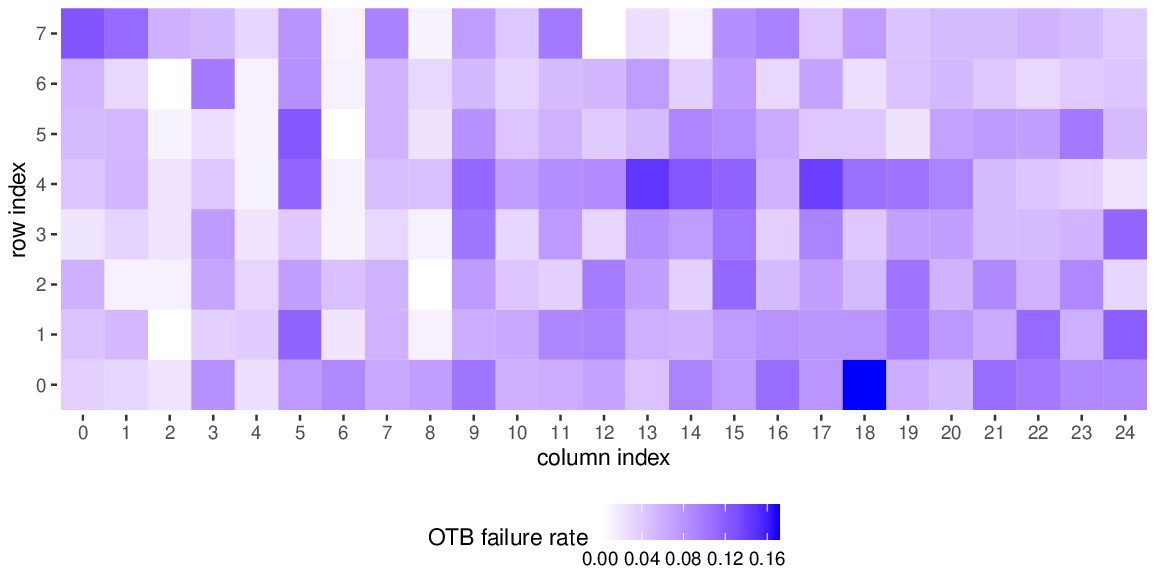}
		\caption{The OTB failure proportion of GPU nodes in different cabinet locations of Titan supercomputer.}\label{fig:gpu.org.failrate}
	\end{figure}

	In the Titan supercomputer, GPU nodes are placed in $8 \times 25$ cabinets. There are 3 cages within each cabinet, 8 slots within each cage, and 4 nodes within each slot. 
	Since \shortciteN{ostrouchov2020gpu} found that GPU failure is related to the heat dissipation in the room, we consider using the row and column of cabinets as the spatial location of each GPU node, as it can serve as a proxy for the regional temperature. The cage, slot, and node positions are considered as covariates. Table~\ref{tbl:gpu.data.illustration} gives an illustration of the information contained in our dataset.  Figure~\ref{fig:gpu.org.failrate} presents the failure proportion of GPU nodes at each cabinet location. It is clear that heterogeneity exists in failure rates at different locations and tends to be similar in adjacent cabinets compared to those far away.

	\begin{table}
		\caption{Illustration of Titan GPU lifetime data structure.}\label{tbl:gpu.data.illustration}
		\centering

		\begin{tabular}{cccccccc}  
			\hline 
			\hline 
			SN & Column & Row & Cage & Slot & Node & Fail Type & Failure Time (years) \\  
			\hline
			323512026070 &  11 &   4 &   0 &   7 &   2 & censor & 5.91 \\  323512026071 &  18 &   6 &   1 &   3 &   0 & censor & 3.68 \\ 
			323512026072 &  24 &   1 &   0 &   4 &   2 & censor & 0.44 \\ 
			323512026108 &   9 &   1 &   0 &   0 &   2 & OTB & 4.00 \\  
			323512026161 &  12 &   4 &   2 &   6 &   2 & OTB & 3.15 \\ 
			\hline
			\hline 
		\end{tabular}
	\end{table}
	
	
	\subsection{Pine Tree Lifetime Data}\label{sec:vi.tree.description}
\begin{figure}
		\centering
		\includegraphics[width = \linewidth]{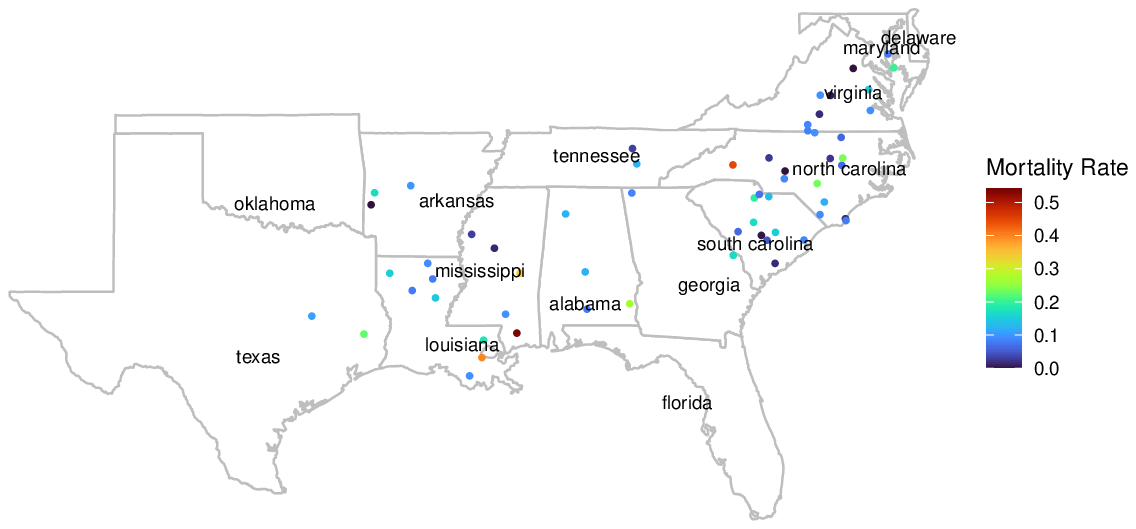}
		\caption{The mortality rate of pine trees across 60 selected US sites. To avoid the overlapping between points, jitters with a uniformly distributed random noise are added to the original site location.}\label{fig:tree.mortality}
	\end{figure}

	The pine tree dataset is from \shortciteN{li2015survival}. This dataset tracks the lifetime of trees under varying living conditions and treatments. A total of 42{,}525 pine trees across 182 sites are collected throughout the Piedmont and Atlantic Coastal Plain regions. During each tree's lifetime, variables such as total height (TH), diameter at breast height (DBH), and crown class are recorded every three years up to 7 times. Other time-invariant variables, such as physical region and thinning treatment (control, light thinning, and heavy thinning) applied to the tree are also collected. In this study, due to the computational limitation, a subset of 60 randomly selected sites with 14{,}044 trees is used for analysis. Table \ref{tbl:tree.variable.exp} presents the exploratory variables and summary statistics of pine trees in the 60 selected sites.  The event of interest is the tree mortality. A tree is recorded as censored if it survives till the $7$th follow-up period. Figure~\ref{fig:tree.mortality} presents the mortality rate of pine trees across 60 selected sites. To avoid overlapping markers in the visualization, we applied small uniform random displacements to the original coordinates.

	\begin{table}
		\caption{The description and summary statistics of explanatory variables of 60 selected sites in the pine tree dataset.}\label{tbl:tree.variable.exp}
		\begin{center}
			
			\begin{tabular}{llc}
				\hline
				\hline
				Variable & Description & $N$ (\%)\\
				\hline
				SN & A serial number assigned to each tree&$14044\ (100\%)$\\
				\cline{2-3}
				Thinning Intensity &1-control&$3370{\ (24.00\%)}$\\
				&2-light thinning&$5674{\ (40.40\%)}$\\
				&3-heavy thinning&$5000{\ (35.60\%)}$\\
				\cline{2-3}
				Physical Region &1-Coastal Plain 	&$7448{\ (53.00\%)}$\\
				&2-Piedmont           	&$5261{\ (37.50\%)}$\\
				&3-other types          &$1335{\ (9.50\%)}$\\
				\cline{2-3}
				Crown Class & 1-dominant	&	$5543{\ (39.50\%)}$\\
				& 2-codominant		&	$9603\ (68.40\%)$\\
				& 3-intermediate	&	$3823{\ (27.20\%)}$\\
				& 4-suppressed		&	$1302\ (9.30\%)$\\
				\hline
				\hline
				Variable & Description & Mean (SD)\\
				\hline
				DBH&	Diameter at breast height (cm)&$7.39\ (2.19)$\\
				TH &	Total height of the tree (meters)&$49.36\ (13.86)$\\
				\hline
				\hline
			\end{tabular}
		\end{center}
	\end{table}
	
	\section{Methodologies} \label{sec:methods}
	In Section \ref{sec:spatial.model}, we first introduce the two models, the cumulative exposure model (CEM) and proportional hazard model, that are most appropriate for analyzing the spatially correlated lifetime datasets described in Section~\ref{sec:data_prep}. Then, in Section~\ref{sec:vi}, we present the Bayesian framework for these models and detail the variational inference approach, which incorporates two types of divergences, along with the parameter estimation procedure.
	
	\subsection{Modeling Spatially Correlated Lifetime Data}\label{sec:spatial.model}
	We first introduce some general notations for both models. Suppose there are $m$ distinct locations, $s_1, \dots, s_m$. Let $t_{ij}$ be the time-to-event for the $j$th unit in the $i$th location $s_i$, where $i = 1, \dots, m, j = 1, \dots, n_i$. Here $n_i$ is the number of units in location $s_i$, and the total number of units is $n = \sum_{i=1}^m n_i$. Let $\delta_{ij}$ be the corresponding censoring indicator. Set $\delta_{ij} = 1$ if the unit failed, and otherwise $\delta_{ij} = 0$. Denote $\xvec_{ij}(t)$ to be the $p$-dimensional vector of explanatory variables at time $t$ at location $i$. Here we consider a general notation of the covariates vector $\xvec_{ij}(t)$, which is a function of time $t$. For a time-invariant variable, the corresponding elements of $\xvec_{ij}(t)$ can be set to constants. For categorical variables, a dummy variable coding can be implemented. The time-to-event data collection can be denoted by $\Dcal = \left\{t_{ij}, \delta_{ij}, \xvec_{ij}(t): t\leq t_{ij}, i = 1, \dots, m, j = 1, \dots, n_i\right\}$. 
	\subsubsection{Cumulative Exposure Model}\label{sec:vi.aft}
	
	
	For the GPU dataset, the problem of interest is to model the time-to-event of GPU nodes using their physical architecture covariates and the spatial locations. Accelerated failure time (AFT) model, as a special case of the CEM, is in an additive form that can easily incorporate multiple covariates and random effects. \citeN{newby1988accelerated} and \citeN{meeker2022statistical} demonstrate that AFT can properly model the mechanism of failure. The model form of AFT assumes that the mechanism of failure does not change with the covariates, but is simply accelerated or slowed down. The model's applicability in engineering is further supported by studies such as \citeN{hong2010field}, \citeN{kundu2019weibull}, \shortciteN{jie2022gpu}. Therefore, in this study, the AFT model is considered for the GPU dataset.
	
	For presentation convenience, we first describe the general model, CEM, which can handle both time-varying and time-invariant covariates. We then demonstrate how the CEM can be simplified into an AFT model. The CEM describes the cumulative damage level of the product by a certain time $t$ given time-varying covariates $\xvec(t)$ \cite{hong2013field}. Let $\betavec$ be the coefficient vector of covariates. The cumulative exposure $u_{ij}(t)$ of unit $j$ in location $i$ by time $t$ is defined as
	\begin{equation*}
		u_{ij}(t) = \int_0^t \exp\left[-\xvec_{ij}(s)\tran\betavec\right]ds.
	\end{equation*}
	The unit fails at time $T$ when the amount of cumulative exposure reaches a random threshold $U$. Let $T_{ij}$ be the random time-to-event of the $j$th unit in the $i$th location. Usually, the log cumulative exposure $\log[u_{ij}(T_{ij})]$ is assumed to follow a location-scale distribution. When considering the spatial random effect, the model can be written as
	\begin{equation}\label{eq:cmu.exp.model}
		\log\left[u_{ij}(T_{ij})\right] = \mu + \gamma_i + \sigma \epsilon_{ij},
	\end{equation}
	where $\gamma_{i}$ is the spatial random parameter that represents the random effect of the $i$th location, and $\epsilon_{ij}$ is independent and identically distributed and follows a standard location-scale distribution. Here, $\mu$ and $\sigma$ are the corresponding location and scale parameters. Let $\gammavec = \left(\gamma_1, \dots, \gamma_m\right)\tran$ be the spatial random parameter vector. The random effect vector $\gammavec$ is assumed to follow a multivariate normal distribution $\gammavec \sim \MVN(\zerovec, \Sigma)$, where $\Sigma = s^2_{\gamma}\Omega$. Here, $s^2_{\gamma}$ describes the overall spatial variability and $\Omega = \left(\rho_{i,i'}\right)_{m \times m}$ is the correlation matrix. The $(i,i')$th element of the correlation matrix $\rho_{i,i'}$ describes the spatial correlation between the random effect of location $s_i$ and $s_{i'}$. Commonly used correlation functions include exponential, Gaussian, and Mat\'ern correlations. In this paper, we use the exponential correlation function $\rho_{i,i'} = \exp[-d(s_i, s_{i'})/\nu], \nu > 0,$
	where $d(s_i, s_{i'})$ is the Euclidean distance between locations $s_i$ and $s_{i'}$.

	Denote the model parameter vector as $\thetavec = (\mu, \betavec\tran, \gammavec\tran, \sigma, s_{\gamma}^2, \nu)\tran$. Given the distributional assumptions, the likelihood of the model parameters is
	\begin{align}\label{eq:cum.expo.L}
		\Lcal(\thetavec|\Dcal) &= L(\gammavec, s_{\gamma}^2, \nu) L(\mu, \betavec, \sigma|\Dcal, \gammavec) \\ \nonumber
		&= (2\pi)^{-m/2}|\Sigma|^{-1/2} \exp\left(-\frac{1}{2}\gammavec\tran \Sigma^{-1} \gammavec\right)
		\prod_{ij}\left[\frac{1}{\sigma t_{ij}}\phi(z_{ij})\right]^{\delta_{ij}} \left[1-\Phi(z_{ij})\right]^{1-\delta_{ij}},
	\end{align}
	where
	$$z_{ij} = \frac{\log(u_{ij}) - \gamma_i}{\sigma}.$$
	Here $u_{ij} = u(t_{ij})$ is the cumulative exposure level at the time-to-event $t_{ij}$ of $j$th unit in the $i$th location, and $\phi(\cdot)$ and $\Phi(\cdot)$ are the probability density function (pdf) and cumulative distribution function (cdf) of standard location-scale distribution, respectively.
	
	In the GPU dataset, the covariates are time-invariant. Therefore, the AFT model, a special case of CEM, is considered for analyzing the dataset. When the covariates are time-invariant, the cumulative exposure becomes $$u(t) =\int_0^t \exp\left(-\xvec_{ij}\tran\betavec\right)ds = \exp\left(-\xvec_{ij}\tran\betavec\right)t.$$ The CEM model in (\ref{eq:cmu.exp.model}) can be simplified to an AFT model:
	\begin{equation}\label{eq:aft}
		\log\left[u_{ij}(T_{ij})\right] = -\xvec_{ij}\tran\betavec + \log(T_{ij}) = \mu + \gamma_{i} + \sigma \epsilon_{ij},
	\end{equation}
	which is equivalent to
	$$\log(T_{ij}) = \mu +\xvec_{ij}\tran\betavec + \gamma_{i} + \sigma \epsilon_{ij}.$$
	The $z_{ij}$ term in the likelihood function (\ref{eq:cum.expo.L}) becomes
	$z_{ij} = [\log(t_{ij}) - \xvec_{ij}\tran\betavec - \gamma_i]/{\sigma}.$

	\subsubsection{Proportional Hazard Model with Time-dependent Covariates}\label{sec:vi.ph}
	
	For the pine tree survival data, how the survival rate of pine trees is impacted by the thinning treatment, environmental factors across different regions is of interest. The PH model \cite{cox1972regression} provides a suitable framework for analyzing such lifetime data. This model is well-suited for the pine tree dataset because it accommodates time-varying covariates (e.g., DBH and TH) as well as random effects. Additionally, it allows for direct estimation of pine tree survival rates. The PH model assumes that the covariates as well as random effects are multiplicatively related to the hazard. The hazard function of the $j$th unit in the $i$th location is modeled as
	\begin{equation}\label{eq:ph}
		h_{ij}(t) = h_0(t) \exp\left[\xvec_{ij}(t)\tran\betavec + \gamma_i\right],
	\end{equation}
	where $h_0(t)$ is the baseline hazard function, $\betavec$ is the coefficient vector for covariates, and $\gamma_{i}$ is the spatial random effect at location $s_i$. Then the cumulative hazard function can be expressed as:
	\begin{equation*}
		H_{ij}(t) = \int_0^t h_{ij}(s)ds = \exp(\gamma_i)\int_0^t h_0(t) \exp\left[\xvec_{ij}(t)\tran\betavec\right]ds.
	\end{equation*}
	
	The survival function of the $j$th unit in the $i$th location can be obtained by $S_{ij}(t) = \exp\left[-H_{ij}(t)\right]$. In this paper, to accommodate the Bayesian framework, we consider a parametric form of the baseline hazard $h_0(t; \thetavec_h)$, where $\thetavec_h$ is the parameter vector. Commonly used hazard functions, like Weibull or log-normal hazard function can be considered. Then the model parameter can be written as $\thetavec = \left(\betavec\tran, \gammavec\tran, \thetavec_h\tran, s_{\gamma}^2, \nu\right)\tran.$ The likelihood of the model parameter is
	\begin{align*}
		L(\thetavec|\Dcal) &= L(\gammavec, s_{\gamma}^2, \nu) L( \betavec, \thetavec_h|\Dcal, \gammavec) \\
		&= (2\pi)^{-m/2}|\Sigma|^{-1/2} \exp\left(-\frac{1}{2}\gammavec\tran \Sigma^{-1} \gammavec\right)
		\prod_{ij}\left[h_{ij}(t_{ij})\right]^{\delta_{ij}} S_{ij}(t_{ij}).
	\end{align*}
	
	
	\subsection{ Variational Inference for Spatially Correlated Lifetime Data} \label{sec:vi}
	
	For statistical inferences of the model parameter $\thetavec$, priors are assigned to $\thetavec$ under the Bayesian framework. In the above models, since the scale parameter $s_{\gamma}^2$ and length-scale parameter $\nu$ in the spatial covariance matrix are positive, inverse-gamma priors are considered in this study. Non-informative priors are considered for the covariate parameter vector $\betavec$. The parameters $\mu$ and $\sigma_l = \log(\sigma)$  in the spatial AFT model and $\thetavec_h$ in the spatial PH model are also assumed to follow non-informative priors. The Bayesian framework for the spatial lifetime model can be expressed as:
	\begin{align*}
		\gammavec & \propto \MVN(\zerovec, s_{\gamma}^2\Omega),\\
		s_{\gamma}^2 &\propto \IGAM(a_{\sigma}, b_{\sigma}),\\
		\nu &\propto \IGAM(a_{\nu}, b_{\nu}),\\
		\betavec_p &\propto \onevec_p.\\
		\shortintertext{For spatial AFT:}
		\log\left[u_{ij}(T_{ij})\right] &= \mu + \gamma_{i} + \sigma \epsilon_{ij},\\
		\mu & \propto 1,  \quad \sigma_l \propto 1.\\
		\shortintertext{For spatial PH:}
		h_{ij}(t) &= h_0(t;\thetavec_h) \exp\left[\xvec_{ij}(t)\tran\betavec + \gamma_i\right],\\
		\thetavec_h & \propto 1.
	\end{align*}
	Here, $\IGAM$ stands for inverse-gamma distribution. The parameter $a_{\sigma}, b_{\sigma}$ and $a_{\nu}, b_{\nu}$ are the shape and scale parameters for $s_{\gamma}^2$ and $\nu$, respectively.
	
	Denote the priors for $\thetavec$ by $p(\thetavec)$. The posterior distribution is proportional to the joint density function of the data collection $\Dcal$ and model parameters $\thetavec$, which is
	\begin{align}\label{eq:joint.dense.aft}
		p(\thetavec|\Dcal) \propto p(\Dcal|\thetavec)p(\thetavec) = \Lcal(\thetavec|\Dcal)p(\thetavec)
		= \Lcal(\thetavec|\Dcal)p(s^2_{\gamma})p(\nu).
	\end{align}
	
	Instead of using MCMC methods to draw samples from the posterior, in this study, we focus on how to use VI for model inference. The key idea of VI is to use a variational distribution $q(\thetavec|\etavec)$ to approximate the exact posterior. Here $\etavec$ is the parameter vector in the variational probability distribution. One would like to find $\etavec$ that allows the variational distribution $q(\thetavec|\etavec)$ to be as close to the exact posterior $p(\thetavec|\Dcal)$ as possible for better approximation. Usually, divergence metrics are used to evaluate the distance between  $p(\thetavec|\Dcal)$ and $q(\thetavec|\etavec)$, and the optimal $\etavec$ is obtained by minimizing the distribution distance. Given the estimated $\etavec$, one can sample model parameters from the variational distribution $q(\thetavec|\etavec)$ and conduct inferences as in MCMC methods. Note that VI reformulates the Bayesian inference problem into an optimization problem. So the computational cost of VI depends on the complexity of the optimization problem. Often the case, the variational distribution $q(\thetavec|\etavec)$ is chosen to be relatively simple to gain the computational advantages. While this simplification enables faster computation compared to MCMC, it may introduce approximation bias since the variational family may not fully capture the true posterior. Nevertheless, for spatially correlated lifetime data, VI remains valuable for efficient explorations of complex models.

	\subsubsection{Two Types of Divergence}
	One important aspect of VI is the choice of the divergence metric used to measure the distance between the true posterior $p(\thetavec|\Dcal)$ and the approximate posterior $q(\thetavec|\etavec)$. KL divergence is the most commonly used metric in VI estimation procedure. However, for some complex model and data structures, the KL divergence may not be flexible enough to capture the true posterior's characteristics. Recent advances in VI have extended the KL divergence to richer divergence families, such as R\'enyi's $\alpha$-divergence, to better approximate the exact posterior distribution. Therefore, in this study, we are interested in the inference performance of both KL divergence and $\alpha$-divergence.
	
	The $\KL$ divergence between $p(\thetavec|\Dcal)$ and $q(\thetavec|\etavec)$ is defined as:
	\begin{align}\label{eq:kl}
		\KL\left[q(\thetavec|\etavec)||p(\thetavec|\Dcal)\right]
		&= \int q(\thetavec|\etavec) \log\left[ \frac{ q(\thetavec|\etavec)}{p(\thetavec|\Dcal)} \right]d\thetavec \\
		&=  \EE_{q(\thetavec|\etavec)} \{\log[q(\thetavec|\etavec)]\} - \EE_{q(\thetavec|\etavec)}\{\log\left([p(\Dcal, \thetavec)]\right)\} + \log[p(\Dcal)].\nonumber
	\end{align}
	
	Because $\log[p(\Dcal)]$ is intractable, we can not directly minimize KL divergence. An equivalent optimization problem is to maximize the evidence lower bound (ELBO):
	\begin{equation}\label{eq:elbo}
		\Lcal_{\text{VI}} = \log[p(\Dcal)] - \KL(q(\thetavec|\etavec)||p(\thetavec|\Dcal)) = \EE_{q(\thetavec|\etavec)}\left\{\log\left[\frac{p(\thetavec, \Dcal)}{q(\thetavec|\etavec)}\right] \right\}.
	\end{equation}
	
	One characteristic of the KL divergence is that it is an asymmetric measurement. That means $\KL\left[q(\thetavec|\etavec)||p(\thetavec|\Dcal)\right] \neq \KL\left[p(\thetavec|\Dcal)||q(\thetavec|\etavec)\right]$. The choice of KL divergence direction affects the behavior of the variational distribution. 
	
	R\'enyi's $\alpha$-divergence is a more general metric that measures the closeness of $p(\thetavec|\Dcal)$ and $q(\thetavec|\etavec)$. It is defined as
	\begin{equation*}
		D\left[q(\thetavec|\etavec)||p(\thetavec|\Dcal)\right] =
		\frac{1}{\alpha - 1} \log \left[\int q(\thetavec|\etavec)^{\alpha} p(\thetavec|\Dcal)^{1-\alpha} d\thetavec\right],
	\end{equation*}
	where $\alpha > 0, \alpha \neq 1$. When $\alpha \rightarrow 1$, the $\alpha$-divergence approaches the KL divergence. So the $\alpha$-divergence can be considered as the root family of KL divergence.
	
	Similar to the KL divergence, we can not compute the $\alpha$-divergence directly. Alternatively, we can maximize the variational R\'enyi (VR) bound, which is defined as
	\begin{equation}\label{eq:ad.obj}
		\Lcal_{\alpha} = \log[p(\Dcal)] - D\left[q(\thetavec|\etavec)||p(\thetavec|\Dcal)\right]
		=\frac{1}{1-\alpha } \log \left\{\EE_{q(\thetavec|\etavec)}\left[\left(\frac{p(\thetavec, \Dcal)}{q(\thetavec|\etavec)}\right)^{1-\alpha} \right] \right\}.
	\end{equation}
	
	Unlike the original R\'enyi's $\alpha$-divergence's positive support, the VR lower bound can be extended to $\alpha \leq 0$. Besides, the objective $\Lcal_{\alpha}$ is continuous and non-increasing on $\alpha$ \cite{li2016renyi}. When $\alpha \rightarrow -\infty$, it encourages mass-covering behavior. That means $\thetavec$ is encouraged to explore a wider range of possible values. When $\alpha \rightarrow \infty$, it encourages the zero-forcing behavior. That means the approximation tends to return an over-confident $\thetavec$ estimate with a small variance. By adjusting the $\alpha$ value, we can encourage flexible variational distribution.
	
	\subsubsection{Black-Box VI Parameter Estimation Procedure}\label{sec:vi.algorithm}
	
	For some specific models with convenient conditional distributions, the VI with KL divergence parameter estimation can be obtained via a closed-form coordinate ascent algorithm \cite{blackboxvi}. For generic models and $\alpha$-divergence, the VR bound $\Lcal_{\alpha}$ is often intractable due to the model complexity and high-dimensional integration over $\thetavec$. Therefore, we adopt the black-box VI idea as presented in \shortciteN{hernandez2016black} and \citeN{blackboxvi}, and use a Monte Carlo (MC) method to evaluate the VR bound. Since KL divergence is a special case of $\alpha$-divergence with $\alpha \rightarrow 1$, in the following, we focus on the VR bound $\Lcal_{\alpha}$. Specifically, we randomly sample from the variational distribution $\thetavec_k \sim q(\thetavec|\etavec), k = 1, \dots, h$, and $\Lcal_{\alpha}$ in Equation~\ref{eq:ad.obj} can be approximated by using the sample mean to substitute the expectation: 
	\begin{equation}\label{eq:mc.ad.obj}
		\widehat{\Lcal}_{\alpha} = \frac{1}{1-\alpha } \log \left\{\frac{1}{h}\left[\sum_{k=1}^h \left(\frac{p(\thetavec_k, \Dcal)}{q(\thetavec_k|\etavec)}\right)^{1-\alpha} \right]\right\}.
	\end{equation}
	
	Note that Equation~\ref{eq:mc.ad.obj} is a general expression that works for both KL divergence and $\alpha$-divergence. After obtaining the approximated VR bound, $\widehat{\Lcal}_{\alpha}$ can be optimized and updated iteratively to obtain variational parameter estimates $\etavec^*$. We optimize $\widehat{\Lcal}_{\alpha}$ using the Adam algorithm, which is an extension to the stochastic gradient descent method \cite{kingma2014adam}. The derivation for the gradient of the VR bound is shown below:

	\begin{align*}
		\frac{\partial \Lcal_{\alpha}}{\partial \etavec} &= \frac{\partial \Lcal_{\alpha}}{\partial \log[q(\thetavec|\etavec)]} \frac{\partial \log[q(\thetavec|\etavec)]}{\partial \etavec} \\
		&= \left(\frac{1}{1-\alpha} \frac{\partial}{\partial \log[q(\thetavec|\etavec)]} \log\left\{ \int q(\thetavec|\etavec) \left[\frac{p(\thetavec,\Dcal)} {q(\thetavec|\etavec)}\right]^{1-\alpha}d\thetavec\right\}\right) \frac{\partial \log[q(\thetavec|\etavec)]}{\partial \etavec} \\
		&=  \frac{\alpha}{1-\alpha} \frac{ \int q(\thetavec|\etavec) \left[\dfrac{p(\thetavec,\Dcal)} {q(\thetavec|\etavec)}\right]^{1-\alpha} \dfrac{\partial \log[q(\thetavec|\etavec)]}{\partial \etavec}d\thetavec}{ \int q(\thetavec|\etavec) \left[\dfrac{p(\thetavec,\Dcal)} {q(\thetavec|\etavec)}\right]^{1-\alpha}d\thetavec}.
	\end{align*}
	
	Algorithm~\ref{alg:bbvi} describes a general approach to obtain $\etavec^*$. 	
	After getting the variational parameter estimate $\etavec^*$, we can sample model parameters from $\thetavec^l \sim q(\thetavec|\etavec^*), l = 1\dots, L$, and use the sample mean $\sum_{l=1}^L{\thetavec^l}/L$ as the point estimates. The $100(1-\alpha)\%$ credible intervals (CIs) can be obtained from the $\alpha/2$ and $(1-\alpha/2)$ quantiles.
	\begin{algorithm}
		\begin{algorithmic}[1]
			\caption{Obtain parameter estimation through black-box VI}\label{alg:bbvi}
			
			\Require  $q(\thetavec|\etavec)$: the variational distribution and $\tau:$ the stopping threshold.
			\Require $\etavec_0$: initial variational parameter vector.
			\Statex $r \gets 1$ (initialize iteration number);
			\Statex $\etavec \gets \etavec_0$ (initialize variational parameter vector);
			\Statex $\text{stop} = 0$ (initialize the stopping indicator);
			\While{$\text{stop} = 0$}
			\item Take $h$ samples from the variational distribution $\thetavec_k \sim q(\thetavec|\etavec_{r-1}), k = 1, \dots, h$ and compute a stochastic estimate of $\widehat{\Lcal}_{\alpha}$ as in  (\ref{eq:mc.ad.obj}).
			\item Take a gradient descent step in Adam algorithm to obtain the updated parameter estimate $\etavec_{r}$;
			
			\If {$r > 2$ \& ($|\widehat{\Lcal}_{\alpha}^{r} - \widehat{\Lcal}_{\alpha}^{r-1}| /|\widehat{\Lcal}_{\alpha}^{r}|< \tau$ \& $ |\widehat{\Lcal}_{\alpha}^{r-1} - \widehat{\Lcal}_{\alpha}^{r-2}| / |\widehat{\Lcal}_{\alpha}^{r-1}|< \tau$)}
			\State $\text{stop} = 1$
			\EndIf
			\item  $r \gets  r+1$
			\EndWhile
			\item 	\Return $\etavec^* \gets \etavec_{r}$
		\end{algorithmic}
	\end{algorithm}

	\section{Numerical Studies} \label{sec:casestudy}
	
	In this section, we investigate the inference performance of VI using both KL divergence and $\alpha$-divergence on the GPU and pine tree datasets. Section~\ref{sec:hmc} first introduces the benchmark method, HMC, used for comparison. Then the computational efficiency and inference capability of VI compared to HMC are evaluated with two case studies in Sections~\ref{sec:casegpu} and~\ref{sec:casetree}, respectively.
	\subsection{The Benchmark Method: HMC}\label{sec:hmc}
	
	As mentioned in Section~\ref{sec:vi.introuduction}, MCMC is the most often used method for drawing samples from the posterior distribution \cite{li2014application}. Among various MCMC techniques, HMC utilizes Hamiltonian dynamics to rapidly traverse the parameter space and enhance MCMC mixing \cite{neal2011hmc}. As an improvement on HMC, NUTS eliminates the manual parameter tuning step and is extensively applied in Bayesian model inference, including \shortciteN{jie2022gpu}. Thus, NUTS in Stan \shortcite{carpenter2017stan} is considered as a benchmark inference method in this paper. When implementing the log-likelihood functions, positive model parameters are transformed to the range of $(-\infty, \infty)$ and the random effect design matrix is transformed to avoid co-linearity as presented in \shortciteN{jie2022gpu}. The chain length and tree depth parameters can be determined by running short chains and checking convergence first. 
	
	In the case studies, we compare HMC, VI with KL divergence, and VI with $\alpha$-divergence using the following three metrics: the Cox-Snell residual probability plot, the negative log-likelihood (NLL) value, and the computing time. The first two metrics are used to evaluate the inference capability of VI compared to HMC, while the computing time is used to evaluate the computational efficiency. 
	\subsection{Case Study 1: The Titan GPU Lifetime Data}\label{sec:casegpu}
	
	\subsubsection{Bayesian AFT Model for GPU Dataset}
	
	In the GPU dataset, we mainly focused on a more “global” room temperature between cabinets that causes the spatial correlations. Within each cabinet, the temperature is assumed to be relatively stable compared to variations observed across cabinets that are farther apart.  Although researchers from the Oak Ridge National Laboratory found that GPU located at lower cage level may have relatively higher survival probability due to the cooling airflow from bottom to top \shortcite{ostrouchov2020gpu}, in this study, instead of directly modeling the impact of within-cabinet correlation, we modeled the impact of temperature across cage levels as a fixed effect. That is, we considered the cage level as a categorical variable in the model. 
	Similarly, slot and node are also considered as categorical variables. Consider a dummy variable encoding, then there are 12 dummy variable covariates. Denote covariate as $\xvec = \left(1, \xvec_{\text{cage}}\tran, \xvec_{\text{slot}}\tran, \xvec_{\text{node}}\tran\right)\tran$, where $\xvec_{\text{cage}}$ represents the vector of dummy variable encoding of cage variable and similar for $\xvec_{\text{slot}}$ and $\xvec_{\text{node}}$. The spatial AFT model of the $j$th GPU in the $i$th location can be written as
	$$\log(T_{ij}) = \xvec_{ij}\tran\betavec + \gammavec_{i} + \sigma \epsilon_{ij}.$$
	
	For the GPU data, we assume the error term $\epsilon_{ij}$ follows the standard smallest extreme value (SEV) distribution. According to Section~\ref{sec:vi.aft}, the spatial random effect $\gammavec$ is assumed to follow a multivariate normal distribution $\gammavec \sim \MVN(\zerovec, \Sigma)$, where $\Sigma = s^2_{\gamma}\Omega$ and $(\Omega)_{ii'} = \left(\exp[-d(s_i, s_{i'})/\nu]\right)$. Since the exact physical locations of each cabinet in the server room are unavailable, we use the row and column indices as proxies for spatial location coordinates. To have a stable estimation of $\nu$, we standardized the distance along the row and column direction, respectively, into the range [0,1] and then computed the Euclidean distance as the distance between two cabinets. This approach may not provide a perfect approximation of the true spatial positions, nevertheless, it offers a practical way to leverage the available dataset to construct spatial structures. 
	
	Denote the model parameter as $\thetavec = \left(\betavec\tran, \gammavec\tran, \sigma_l, s_{\gamma}^2, \nu\right)\tran$. Then the log posterior can be derived as:
	\begin{align*}
		\log\left[p(\thetavec|\Dcal)\right] =  &\text{Constant}+ \sum_{ij} \left(\delta_i \left\{-\log(t_{ij}) - \sigma_l + \log\left[\phi(z_{ij})\right]    \right\} + (1- \delta_i) \log\left[1-\Phi(z_{ij})\right]        \right) \\& - \frac{1}{2}\left[\log(s_{\gamma}^2 \Omega)  + \frac{\gammavec\tran \Omega^{-1} \gammavec}{s_{\gamma}^2}\right]
		- \frac{1}{2} (a_{\sigma} + 1)\log(s_{\gamma}^2) \\
		&- \frac{b_{\sigma}}{s_{\gamma}^2}
		- \frac{1}{2} (a_{\nu} + 1)\log(\nu) - \frac{b_{\nu}}{\nu}.
	\end{align*}
	
	In the GPU dataset, for the hyper-parameters inside the inverse-gamma prior, we set a weak prior for $s_{\gamma}^2$ by letting $a_{\sigma} = 1$, $b_{\sigma} = 1$ and a strong prior for $\nu$ by setting $a_{\nu} = 13$, $b_{\nu} = 0.1$.
	
	For the variation inference of model parameters, we assume the following variational distributional assumptions:
	\begin{align*}
		\betavec &\sim \MVN(\muvec_{\betavec}, \Sigma_{\betavec}),
		\text{where } \Sigma_{\betavec} = \diag(\sigmavec_{\betavec}^2),\\
		\gammavec & \sim \MVN(\muvec_{\gammavec}, \Sigma_{\gammavec}),
		\text{where } \Sigma_{\gammavec} = \diag(\sigmavec_{\gammavec}^2),\\
		\sigma_l & \sim \N(\mu_{\sigma}, \sigma_{\sigma}^2),\\
		s_{\gamma}^2 &\propto \IGAM(c_{\sigma}, d_{\sigma}),\\
		\nu &\propto \IGAM(c_{\nu}, d_{\nu}),
	\end{align*}
	where $\Sigma_{\betavec}$ and $\Sigma_{\gammavec}$ are diagonal matrices with diagonal elements being $\sigmavec_{\betavec}^2$ and $\sigmavec_{\gammavec}^2$. The variational distributions are independent of each other. In the above framework, $c_{\sigma}, d_{\sigma}$ and $c_{\nu}, d_{\nu}$ are the scale and shape parameters in inverse-gamma distribution for $s_{\gamma}^2$ and $\nu$, respectively. Let $\sigmavec_{l\betavec} = \log(\sigmavec_{\betavec})$, $\sigmavec_{l\gammavec} = \log(\sigmavec_{\gammavec})$ and $\sigma_{l\sigma} = \log(\sigma_{\sigma})$. Then the variational parameter vector can be denoted as $\etavec = \left(\muvec_{\betavec}\tran, \sigmavec_{l\betavec}\tran, \muvec_{\gammavec}\tran, \sigmavec_{l\gammavec}\tran,\mu_{\sigma}, \sigma_{l\sigma}, c_{\sigma}, d_{\sigma},c_{\nu}, d_{\nu} \right)\tran$. The log density of the joint variational distribution is
	\begin{align*}
		\log\left[q(\thetavec|\etavec)\right] =&  \text{Constant}- \frac{1}{2} \sum_{j} \left[2\sigma_{l\betavec,j} + \frac{(\beta_j - \mu_{\betavec,j})^2}{\exp(2\sigma_{l\betavec,j})}\right] \\
		&- \frac{1}{2} \sum_{j} \left[2\sigma_{l\gammavec,j} + \frac{(\gamma_j - \mu_{\gammavec,j})^2}{\exp(2\sigma_{l\gammavec,j})}\right]
		- \frac{1}{2} \left[2\sigma_{l\sigma} + \frac{(\sigma_l - \mu_{\sigma})^2}{\exp(2\sigma_{l\sigma})}\right] \\
		&+ c_{\sigma} \log(d_{\sigma}) - \log[\Gamma(c_{\sigma})] - (c_{\sigma} + 1)\log(s_{\gamma}^2) - \frac{d_{\sigma}}{s_{\gamma}^2} \\
		&+ c_{\nu} \log(d_{\nu}) - \log[\Gamma(c_{\nu})] - (c_{\nu} + 1)\log(\nu) - \frac{d_{\nu}}{\nu},
	\end{align*}
	where $\mu_{\betavec,j}$ is the $j$th element of mean vector $\muvec_{\betavec}$ and $\sigma_{l\betavec,j}$ is the $j$th element of the log diagonal variance vector $\sigmavec_{\betavec}$ for $\betavec$. Similarly for $\mu_{\gammavec,j}$ and $\sigma_{l\gammavec,j}$. And $\Gamma(\cdot)$ denotes the gamma function.
	Plugging in $\log\left[p(\thetavec, \Dcal)\right]$ and $\log\left[q(\thetavec|\etavec)\right]$ in (\ref{eq:elbo}) and (\ref{eq:ad.obj}) we can obtain the objective functions $\Lcal_{\alpha}$ and $\Lcal_{\text{VI}}$ in variational inference. The procedures in Section~\ref{sec:vi.algorithm} can be used to obtain estimates $\etavec^*$. Then we can sample $\thetavec$ from $q(\thetavec|\etavec^*)$ and conduct inferences for the spatial AFT model.
	\subsubsection{Analysis Results}\label{sec:vi.gpu.res}
	For the GPU data, we compare HMC, VI with KL divergence, and VI with $\alpha$-divergence with $\alpha = 0.8$. To understand how well the posterior samples from each inference approach fit the model, we use the residual plot to conduct diagnostics. The censored Cox-Snell residual of the spatial AFT model is an extension of the standardized residual, which is defined as
	$$\widehat{\epsilon}_{ij} = \exp\left[\frac{\log(t_{ij}) - \xvec_{ij}\tran\widehat{\betavec} - \gamma_i}{\widehat{\sigma}}\right].$$
	
\begin{figure}
		\begin{tabular}{ccc}
			\includegraphics[width=0.31\linewidth]{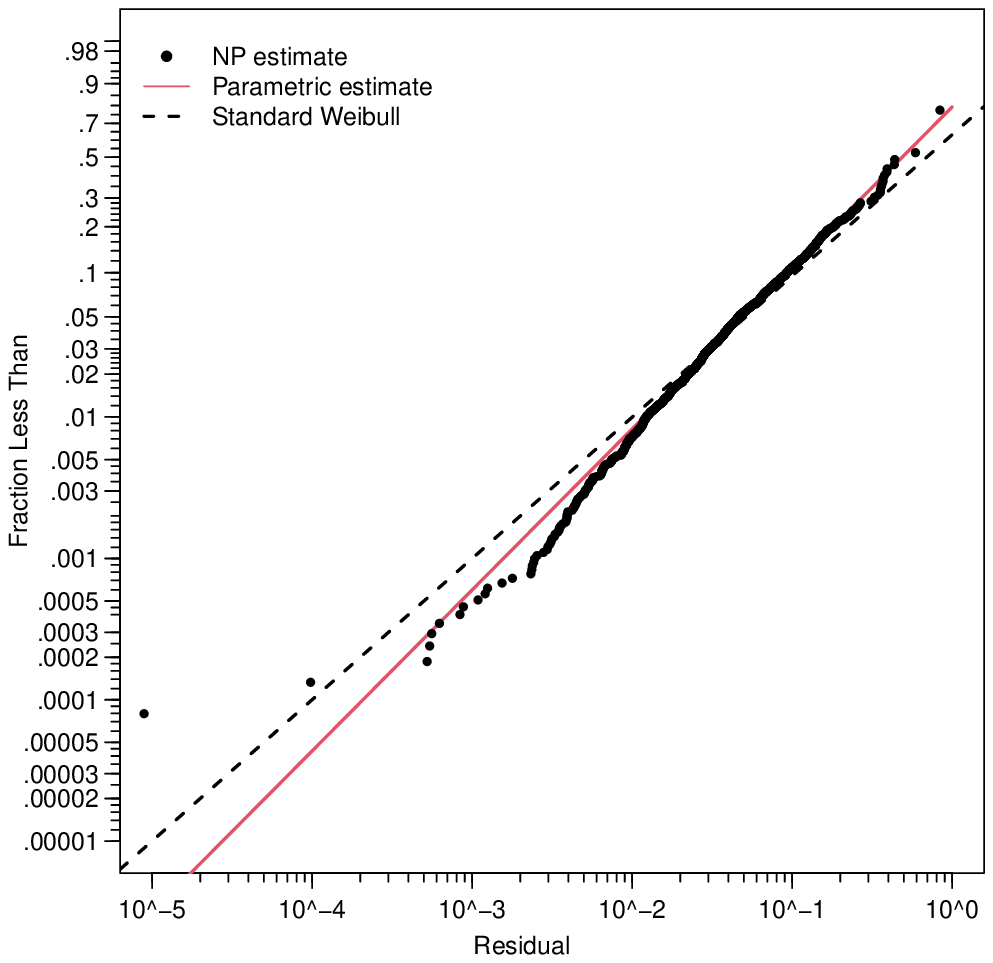} &
			\includegraphics[width=0.31\linewidth]{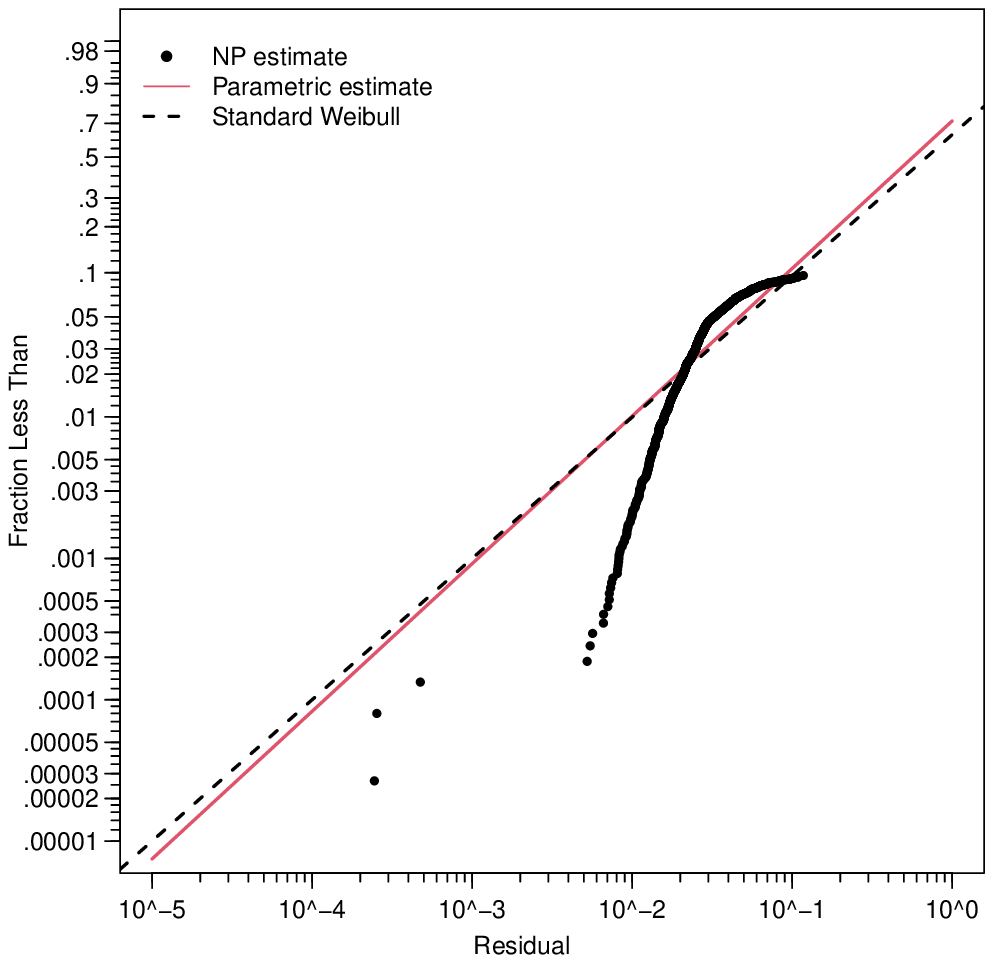} &
			\includegraphics[width=0.31\linewidth]{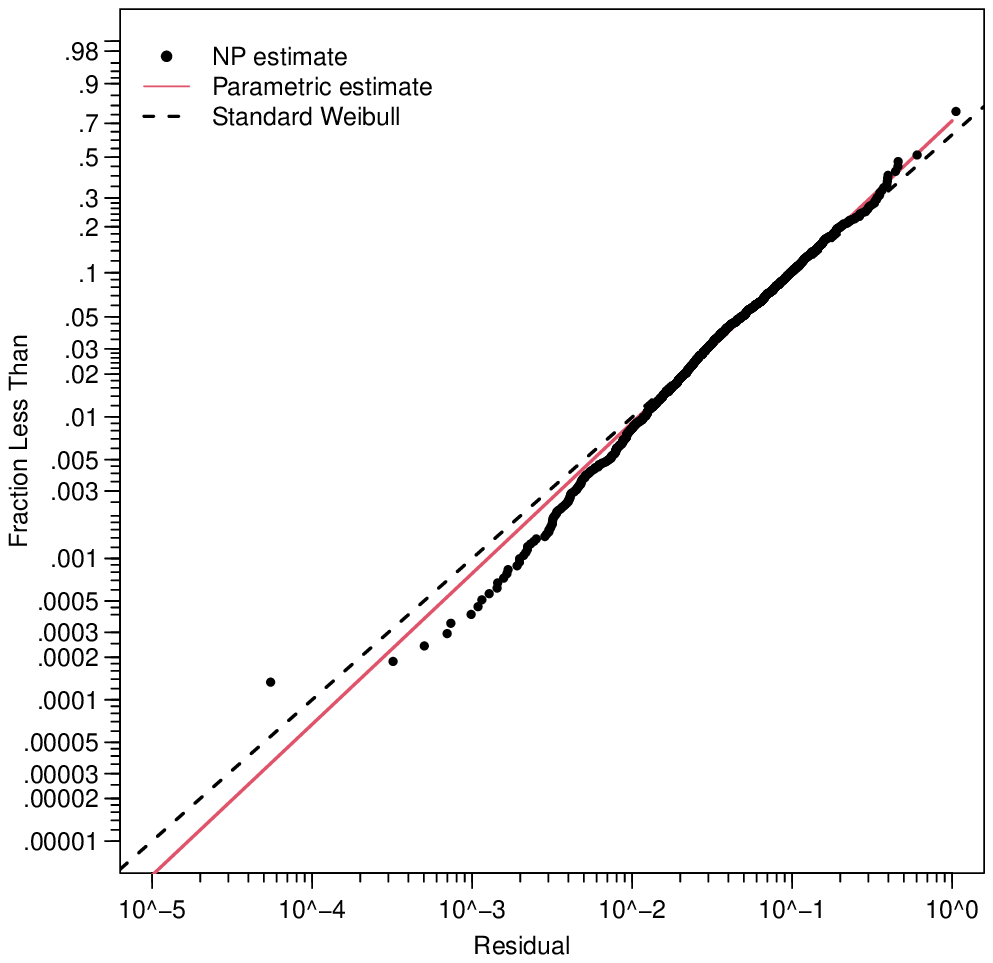}\\
			(a) $\alpha = 0.8$ & (b) KL & (c) HMC
		\end{tabular}
		\caption{Weibull probability plot of residuals for $\alpha$-divergence, KL divergence, and HMC with the GPU data.}\label{fig:res.gpu}
	\end{figure}

	For a censored unit, the corresponding residual is also censored. With the SEV distribution assumption, the residuals should approximately follow a Weibull distribution. The probability plots of censored Cox-Snell residuals of three inference methods are presented in Figure~\ref{fig:res.gpu}. We can see that for $\alpha$-divergence, its probability plot of residual is almost aligned with the standard Weibull distribution. While the KL divergence has some deviation at the tail of residuals.
	
	Table~\ref{tbl:gpu.summary} summarizes the NLL values and computational times across methods. VI with $\alpha$-divergence has an NLL value very close to HMC while using the fastest computation time. Parameter estimates and standard errors (SE) in Table~\ref{tbl:pars.gpu} show that $\alpha$-divergence has estimates closer to HMC benchmarks than KL divergence does. While the Slot effect estimates show slightly greater discrepancies, the absolute value of this effect is small in HMC results, keeping the bias within acceptable limits. From both  $\alpha$-divergence and HMC estimates, we can see that when other variables are fixed, cage 0-1 accelerate a unit's failure compared to the baseline unit with cage 2, while node 0-2 decelerates a unit's failure compared to node 3. The impact of a GPU's slot position is relatively small to its failure.
	
\begin{figure}
		\centering
		\begin{tabular}{cc}
			\includegraphics[width=0.48\linewidth]{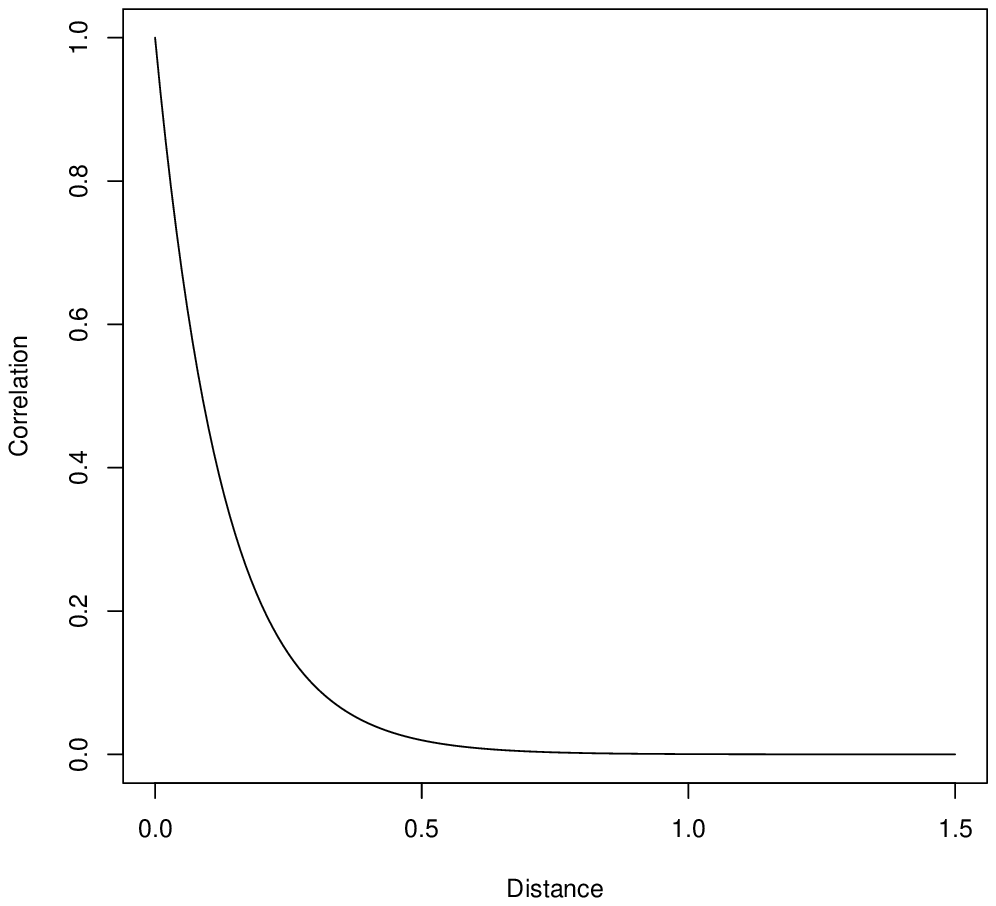}&
			\includegraphics[width=0.48\linewidth]{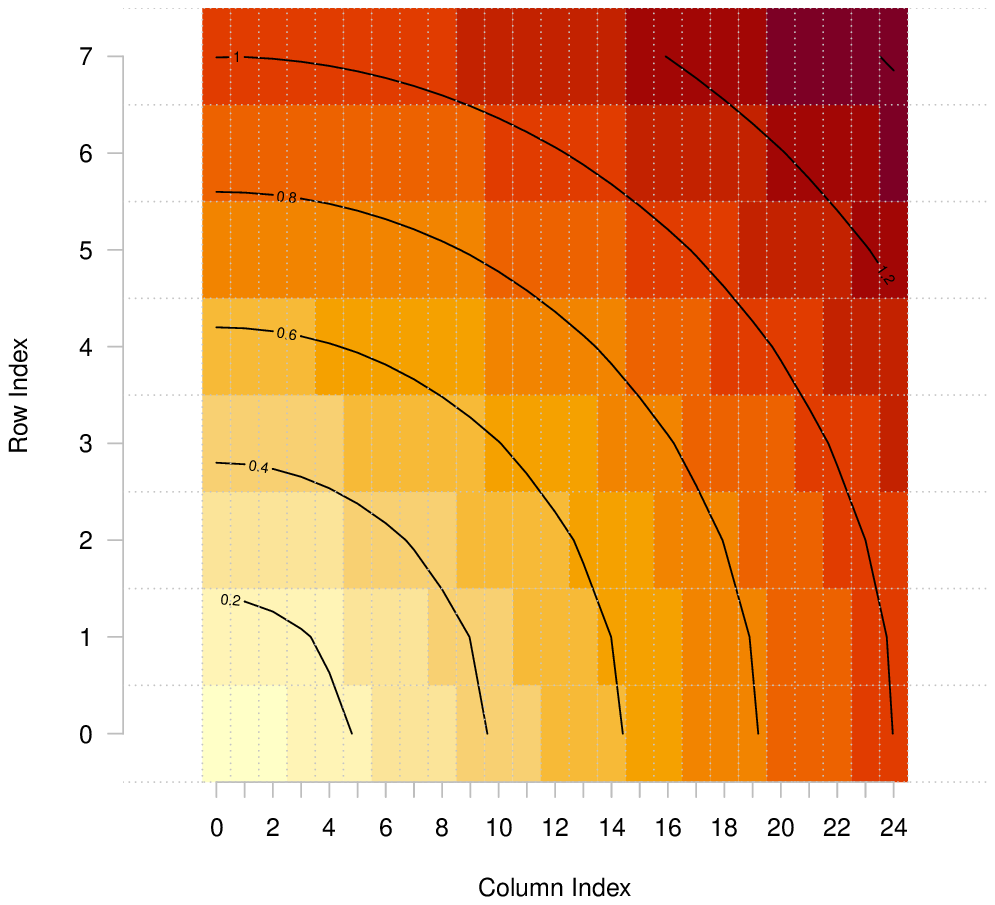} \\
			(a) Spatial Correlation  & (b) Standardized Distance
		\end{tabular}
		
		\caption{The estimated spatial correlation versus distance from VI with $\alpha$-divergence, $\alpha = 0.8$ in the AFT model for the GPU data (a) and the ``standardized grid cell” distance from the cabinet at Row 0, Column 0 to other cabinets (b).}
		\label{fig:gpu.corr}
	\end{figure}

\begin{figure}
		\centering
		\includegraphics[width=0.96\linewidth]{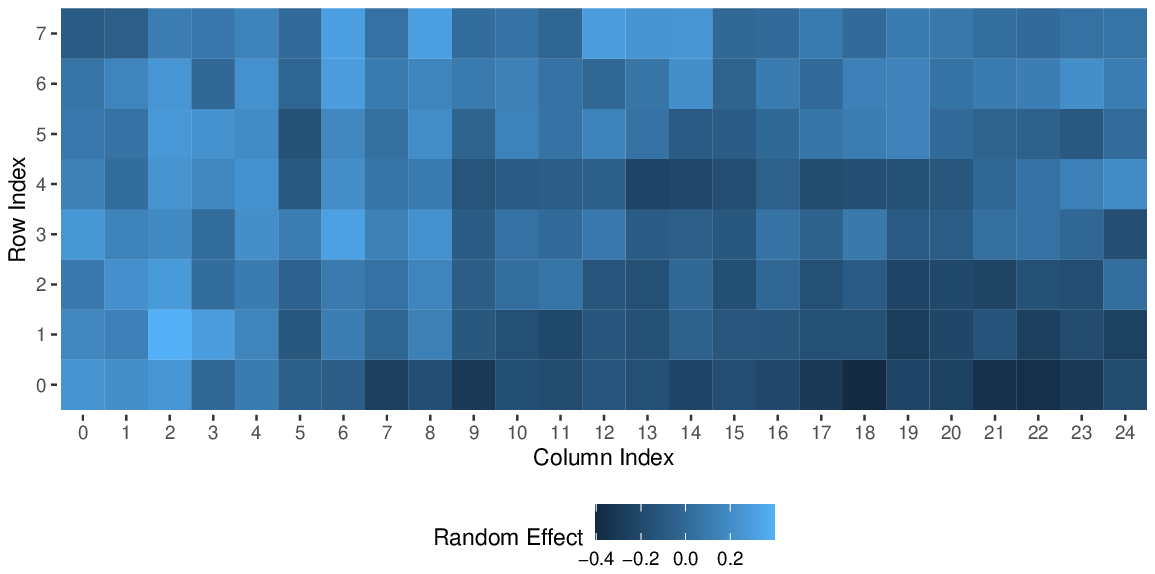} 
		\caption{The heat map of estimated random effects from VI with $\alpha$-divergence, $\alpha = 0.8$ in the AFT model for the GPU data.}
		\label{fig:gpu.rand.heat}
	\end{figure}

	The following results in Figures~\ref{fig:gpu.corr} and~\ref{fig:gpu.rand.heat} present the estimated result obtained using $\alpha$-divergence.  Figure~\ref{fig:gpu.corr}(a) shows the estimated spatial correlation from VI with $\alpha$-divergence as the distance increases. The standardized distance range in the GPU dataset is [0,1.414], with units of ``standardized grid cells" for parameter $\nu$. One ``standardized grid cell” can be difficult to transform back to the original grid cell. For easier interpretation, Figure~\ref{fig:gpu.corr}(b) shows the standardized distances from the cabinet at Row 0, Column 0 to other cabinets. The spatial correlation curve indicates correlations decay to 0.2 at a distance of 0.21, indicating weak correlation. This suggests that cabinets separated by standardized distances greater than 0.21 can be considered weakly correlated or uncorrelated. In original grid cell terms, this corresponds to cabinets separated by more than one row or five columns.
	Figure~\ref{fig:gpu.rand.heat} presents the heat map of the estimated random effect obtained using the VI with $\alpha$-divergence at each location.  The heat map shows that the positions in the bottom right corner tend to decelerate the GPU's failure.  
	
\begin{figure}
		\centering
		\includegraphics[width=0.8\linewidth]{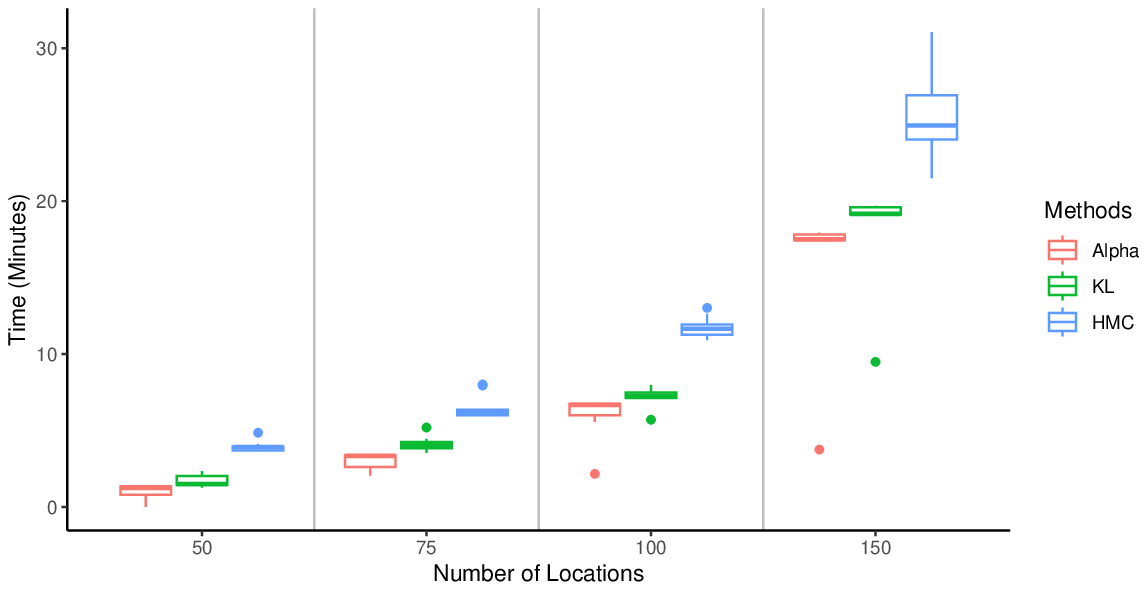}
		\caption{The computing time of three inference methods versus the number of locations with the GPU data.}\label{fig:gpu.time}
	\end{figure}

	It is also interesting to examine whether the number of spatial locations impacts the computing time for each of the three inference methods. A series of subsets with the number of locations $m=50, 75, 100, 150$ are randomly sampled from the original dataset without replacements. The process is repeated 10 times and each of the three methods is applied to the subsets. The boxplot of computing time versus the number of locations is presented in Figure~\ref{fig:gpu.time}. We can see that the computing time difference between HMC and variational approaches increases as the number of locations increases. The computational advantage of $\alpha$-divergence is more obvious when the number of locations is large.

	\begin{table}
		\centering
		\caption{The negative log likelihood and computing time of $\alpha$-divergence, KL divergence, and HMC inferences with the GPU data.}\label{tbl:gpu.summary}
		
		\begin{center}
			\begin{tabular}{rrrr}
				
				\hline
				\hline
				& $\alpha = 0.8$ & KL & HMC \\
				\hline
				NLL & 4481.04 & 5368.06 & 4470.13 \\
				Time (minutes) &42.7& 55.23&  86.53\\
				\hline
				\hline
			\end{tabular}
		\end{center}
	\end{table}
	\begin{table}
		\centering
		\caption{The parameter estimates and SE of three inference methods for the GPU data.}
		\label{tbl:pars.gpu}
		\begin{center}
			
			\begin{tabular}{l|rr|rr|rr}
				\hline
				\hline
				\multirow{2}{*}{Parameter}	& \multicolumn{2}{c}{$\alpha$-divergence} &
				\multicolumn{2}{c}{KL divergence} & \multicolumn{2}{c}{HMC}\\
				\cline{2-7}
				& Estimate & SE & Estimate & SE& Estimate & SE  \\
				\hline
				Constant & 2.0235 & 0.0166 & 2.5989 & 0.0279 & 1.9306 & 0.0278 \\ 
				Cage 0 &0.6922 & 0.0488 & 0.1127 & 0.0483 & 0.6783 & 0.0169 \\ 
				Cage 1 &  0.2755 & 0.0208 & $-$0.0027 & 0.0466 & 0.2734 & 0.0146 \\ 
				Slot 0 & 0.0702 & 0.0284 & 0.0234 & 0.0503 & 0.0412 & 0.0242 \\ 
				Slot 1 & 0.0439 & 0.0337 & 0.0087 & 0.0445 & 0.0285 & 0.0236 \\ 
				Slot 2 & 0.0606 & 0.0348 & $-$0.0221 & 0.0507 & 0.0520 & 0.0240 \\ 
				Slot 3 &0.0675 & 0.0271 & 0.0147 & 0.0484 & 0.0398 & 0.0237 \\ 
				Slot 4 &  0.0811 & 0.0262 & 0.0330 & 0.0486 & 0.0709 & 0.0251 \\
				Slot 5 &0.0141 & 0.0259 & $-$0.0069 & 0.0487 & 0.0119 & 0.0233 \\ 
				Slot 6 & $-$0.0065 & 0.0289 & $-$0.0340 & 0.0496 & $-$0.0083 & 0.0232 \\ 
				Node 0 & $-$0.3049 & 0.0216 & $-$0.0169 & 0.0489 & $-$0.2895 & 0.0192 \\ 
				Node 1 & $-$0.3235 & 0.0231 & $-$0.0187 & 0.0447 & $-$0.3102 & 0.0192 \\ 
				Node 2 & $-$0.0870 & 0.0279 & 0.0169 & 0.0494 & $-$0.0666 & 0.0201 \\ 
				$\sigma$ &  0.2289 & 0.0056 & 0.3866 & 0.0095 & 0.2038 & 0.0053 \\ 
				$s_{\gamma}^2$ & 0.0940 & 0.0318 & 0.0457 & 0.0136 & 0.0488 & 0.0160 \\ 
				$\nu$ &0.1337 & 0.0575 & 1.8981 & 3.0627 & 0.2734 & 0.2426 \\ 
				\hline
			\end{tabular}
			
		\end{center}
	\end{table}

	\subsection{Case Study 2: The Pine Tree Lifetime Data}\label{sec:casetree}
	\subsubsection{Bayesian PH Model for Tree Data}
	
	As introduced in Section~\ref{sec:vi.ph}, we consider the spatial PH model with time-dependent covariates for the pine tree data. For the PH model in Equation~\ref{eq:ph}, the baseline hazard function  $h_0(t)$ needs to be specified. In this study, we consider the Weibull hazard with functional form $h_0(t) = a t^b$ as the baseline function, where $a$ and $b$ are parameters.  Let $\xvec(t) = [\text{DBH}(t), \text{TH}(t), \xvec_{\text{Crown}}(t), \xvec_{\text{PhyReg}}, \xvec_{\text{Trt}}]$, where $ \xvec_{\text{Crown}}(t)$ represent the dummy variable encoding of the crown class at time $t$, similar for $\xvec_{\text{PhyReg}}, \xvec_{\text{Trt}}$. Then the PH model can be expressed as 
	$$h_{ij}(t) = a t^b \exp\left[\xvec_{ij}(t)\tran\betavec + \gamma_i\right].$$
	Similarly, the $\MVN$ distribution is considered for the spatial random effect vector $\gammavec$. In the pine tree dataset, to have stable estimates of the length-scale parameter $\nu$, we used the great circle distance and standardized it into the range [0,1] by dividing the maximum great circle distance among all location pairs.

	Denote the model parameter as $\thetavec = (a_l, b_l, \betavec, \gammavec, s_{\gamma}^2, \nu)$, where $a_l = \log(a), b_l = \log(b)$. The log posterior can be derived as
	\begin{align*}
		\log\left[p(\thetavec|\Dcal)\right] =& \text{Constant} + \sum_{ij} \left\{ \delta_{ij}  \left[\gamma_i + a_l + b_l \log(t) + \xvec_{ij}(t_{ij})\betavec \right] -H_{ij}(t_{ij}) \right\}     \\& - \frac{1}{2}\left[\log(s_{\gamma}^2 \Omega)  + \frac{\gammavec\tran \Omega^{-1} \gammavec}{s_{\gamma}^2}\right]\\
		&
		- \frac{1}{2} (a_{\sigma} + 1)\log(s_{\gamma}^2) - \frac{b_{\sigma}}{s_{\gamma}^2}
		- \frac{1}{2} (a_{\nu} + 1)\log(\nu) - \frac{b_{\nu}}{\nu}.
	\end{align*}
	
	In the pine tree dataset, for the hyper-parameters inside the inverse-gamma prior, we set a weak prior for $s_{\gamma}^2$ by letting $a_{\sigma} = 1$, $b_{\sigma} = 1$ and a strong prior for $\nu$ by setting $a_{\nu} = 5$, $b_{\nu} = 0.1$.We assume the following variational distributional assumptions:
	\begin{align*}
		\betavec &\sim \MVN(\muvec_{\betavec}, \Sigma_{\betavec}),
		\text{where } \Sigma_{\betavec} = \diag(\sigmavec_{\betavec}^2),\\
		\gammavec & \sim \MVN(\muvec_{\gammavec}, \Sigma_{\gammavec}),
		\text{where } \Sigma_{\gammavec} = \diag(\sigmavec_{\gammavec}^2),\\
		a_l & \sim \N(\mu_{a}, \sigma_{a}^2),\\
		b_l & \sim \N(\mu_{b}, \sigma_{b}^2),\\
		s_{\gamma}^2 &\propto \IGAM(c_{\sigma}, d_{\sigma}),\\
		\nu &\propto \IGAM(c_{\nu}, d_{\nu}),
	\end{align*}
	where $\Sigma_{\betavec}$ and $\Sigma_{\gammavec}$ are diagonal matrices with diagonal elements being $\sigmavec_{\betavec}^2$ and $\sigmavec_{\gammavec}^2$. The variational distributions are independent to each other. Let $\sigmavec_{l\betavec} = \log(\sigmavec_{\betavec})$, $\sigmavec_{l\gammavec} = \log(\sigmavec_{\gammavec})$, $\sigma_{la} = \log(\sigma_{a})$ and $\sigma_{lb} = \log(\sigma_{b})$. Then the variational parameter vector can be denoted as $\etavec = \left(\muvec_{\betavec}\tran, \sigmavec_{l\betavec}\tran, \muvec_{\gammavec}\tran, \sigmavec_{l\gammavec}\tran,\mu_{a}, \sigma_{la},\mu_{b}, \sigma_{lb}, c_{\sigma}, d_{\sigma},c_{\nu}, d_{\nu} \right)\tran$. The log density of joint variational distribution is
	\begin{align*}
		\log\left[q(\thetavec|\etavec)\right] =&  \text{Constant}- \frac{1}{2} \sum_{j} \left[2\sigma_{l\betavec,j} + \frac{(\beta_j - \mu_{\betavec,j})^2}{\exp(2\sigma_{l\betavec,j})}\right] \\
		&- \frac{1}{2} \sum_{j} \left[2\sigma_{l\gammavec,j} + \frac{(\gamma_j - \mu_{\gammavec,j})^2}{\exp(2\sigma_{l\gammavec,j})}\right] \\
		& - \frac{1}{2} \left[2\sigma_{la} + \frac{(a_l - \mu_{a})^2}{\exp(2\sigma_{la})}\right]
		- \frac{1}{2} \left[2\sigma_{lb} + \frac{(b_l - \mu_{b})^2}{\exp(2\sigma_{lb})}\right]\\
		&+ c_{\sigma} \log(d_{\sigma}) - \log[\Gamma(c_{\sigma})] - (c_{\sigma} + 1)\log(s_{\gamma}^2) - \frac{d_{\sigma}}{s_{\gamma}^2} \\
		&+ c_{\nu} \log(d_{\nu}) - \log[\Gamma(c_{\nu})] - (c_{\nu} + 1)\log(\nu) - \frac{d_{\nu}}{\nu}.
	\end{align*}
	Similar to Section~\ref{sec:casegpu}, we can apply VI based on the unnormalized posterior $\log\left[p(\thetavec, \Dcal)\right]$ and variational distribution $\log\left[q(\thetavec|\etavec)\right]$ and obtain the estimated $\hat{\thetavec}$.
	\subsubsection{Analysis Results}\label{sec:vi.tree.res}

	For a PH model with time-dependent covariates, the Cox-Snell residual is defined as
	\begin{equation*}
		\widehat{\epsilon}_{ij} = \int_0^{T_{ij}}\widehat{H}_0(t)\exp[\xvec_{ij}(t)\tran\widehat{\betavec} + \widehat{\gamma}_i]dt,
	\end{equation*}
	where $\widehat{H}_0(t)$ is the estimated cumulative baseline hazard rate by plugging in $\widehat{a}$ and $\widehat{b}$. If the model is correct, then $\widehat{\epsilon}_{ij}$ approximately follows an exponential distribution with $\lambda = 1$ and censoring, which is a special case of the Weibull distribution \cite{klein2003survival}. Therefore, we can use Weibull probability plot of residuals to check three inference methods posterior samples.

\begin{figure}
		\begin{tabular}{ccc}
			\includegraphics[width=0.31\linewidth]{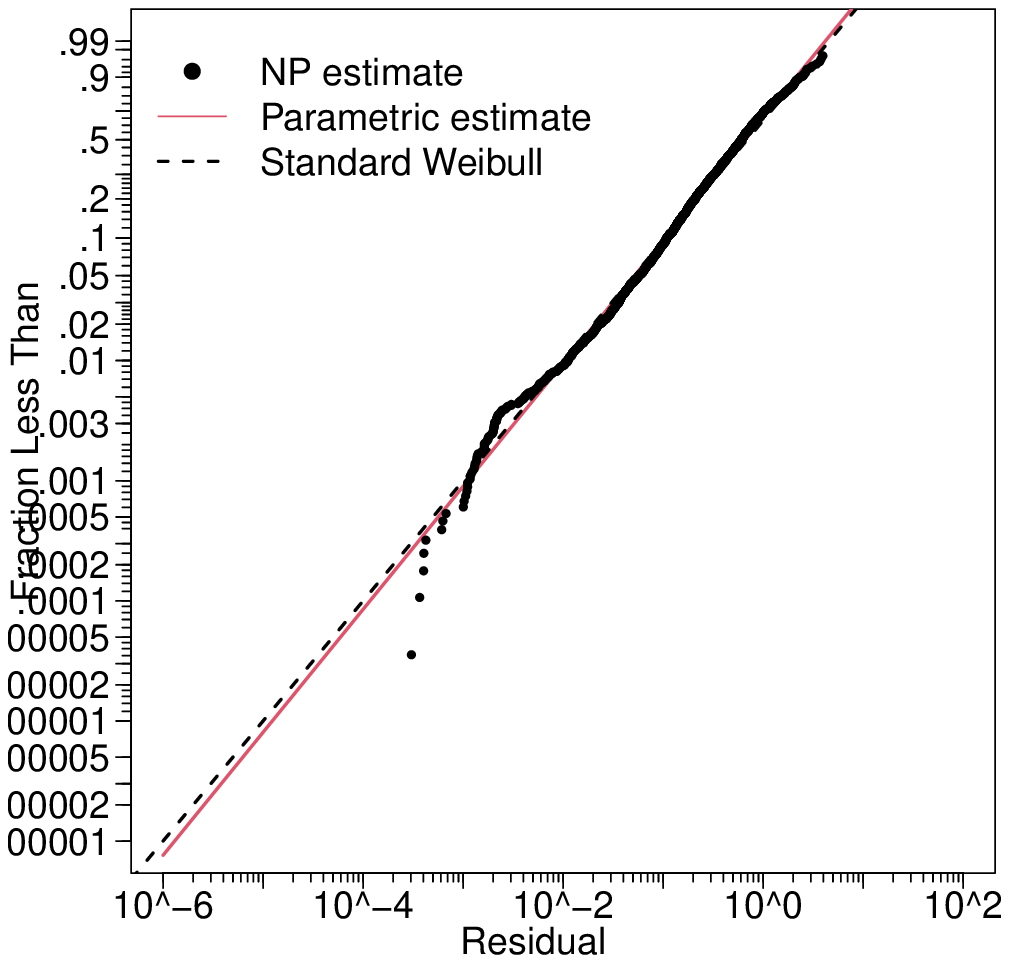} &
			\includegraphics[width=0.31\linewidth]{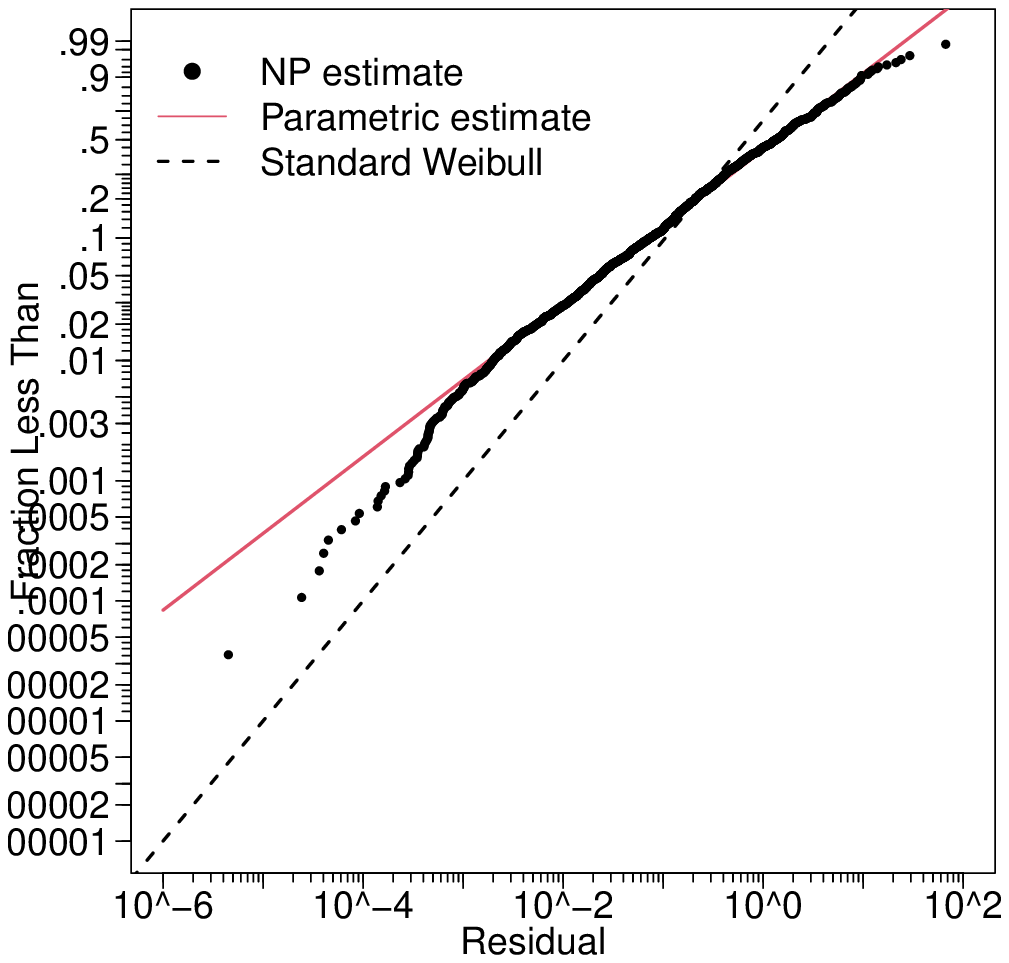} &
			\includegraphics[width=0.31\linewidth]{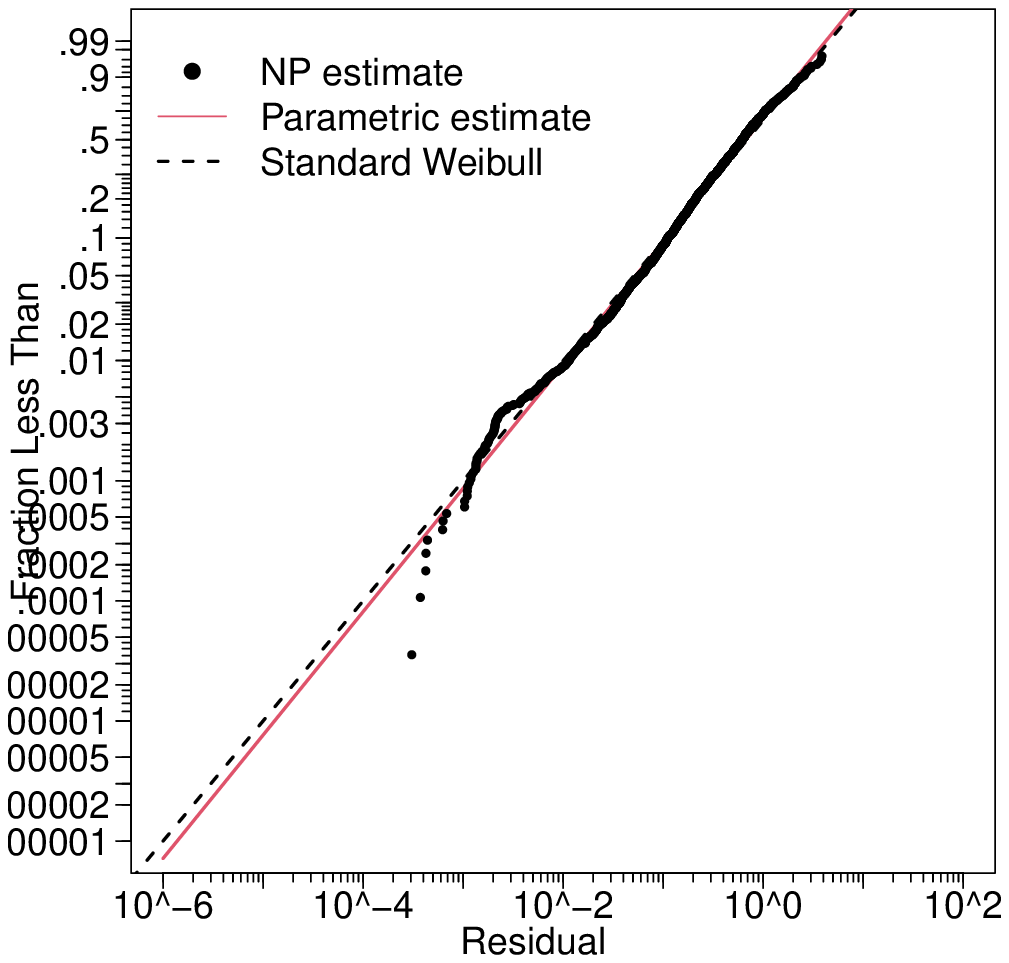}\\
			(a) $\alpha = 0.8$ & (b) KL & (c) HMC
		\end{tabular}
		\caption{Weibull probability plot of residuals for $\alpha$-divergence, KL divergence, and HMC with the pine tree data.}\label{fig:res.tree}
	\end{figure}

	Figure \ref{fig:res.tree} shows the probability plot of residuals for $\alpha$-divergence, KL divergence, and HMC. While there exists curvature at the tail of $\alpha$-divergence residuals, it is close to the HMC method. The KL divergence residuals deviate far more from the standard Weibull distribution. Table~\ref{tbl:tree.summary} summarizes the NLL and computational time of three methods. The NLL value obtained using $\alpha$-divergence is very close to HMC while taking much less computing time. Table ~\ref{tbl:pars.tree} shows the parameter estimates and SE of three inference methods. Compared to the KL divergence, the point estimates of VI with $\alpha$-divergence are closer to HMC, except for the Physical Region effect estimation, while the relative impact of different Physical Regions remains consistent between $\alpha$-divergence and HMC. From both HMC and $\alpha$-divergence estimates, it is seen that if other covariates are fixed, control and light thinning treatment reduce the survival rates compared to
	heavy thinning. Pine trees in the Piedmont region tend to have higher survival rates compared to the Coastal Plain region. Trees with larger DBH and TH tend to have longer lifetime. 
	
\begin{figure}
		\centering
		\begin{tabular}{cc}
			\includegraphics[width=0.49\linewidth]{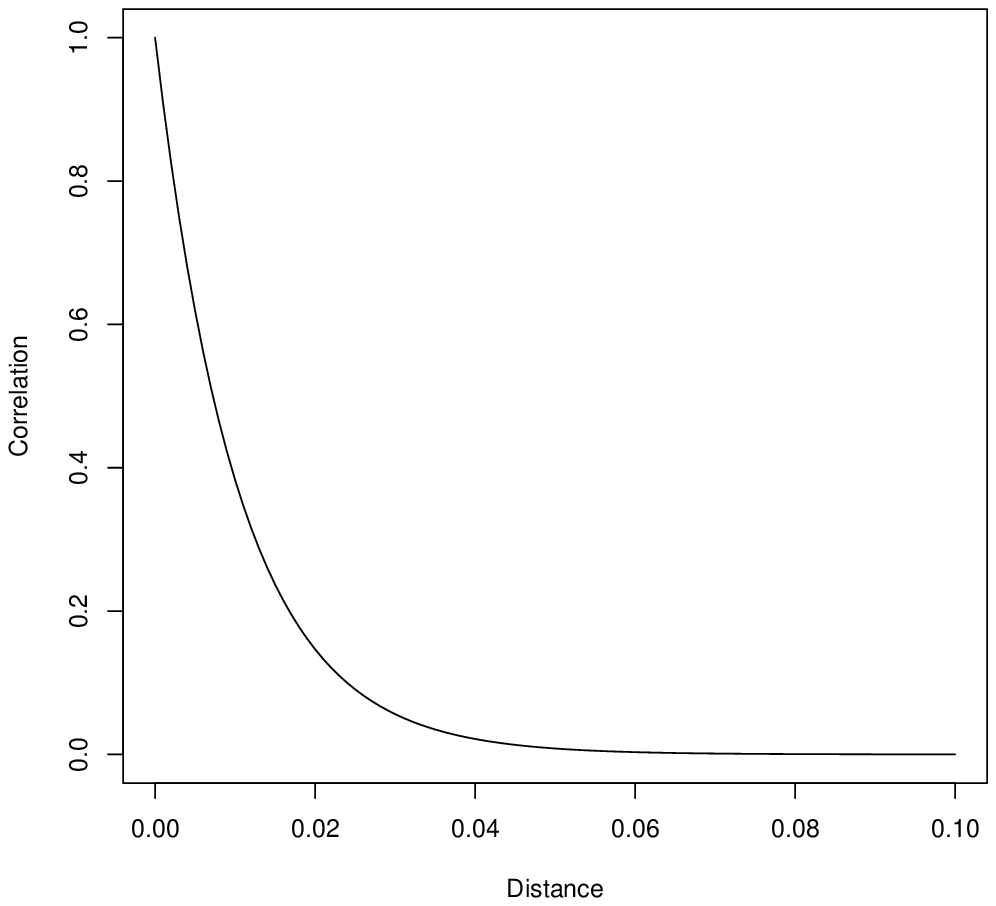} &
			\includegraphics[width=0.49\linewidth]{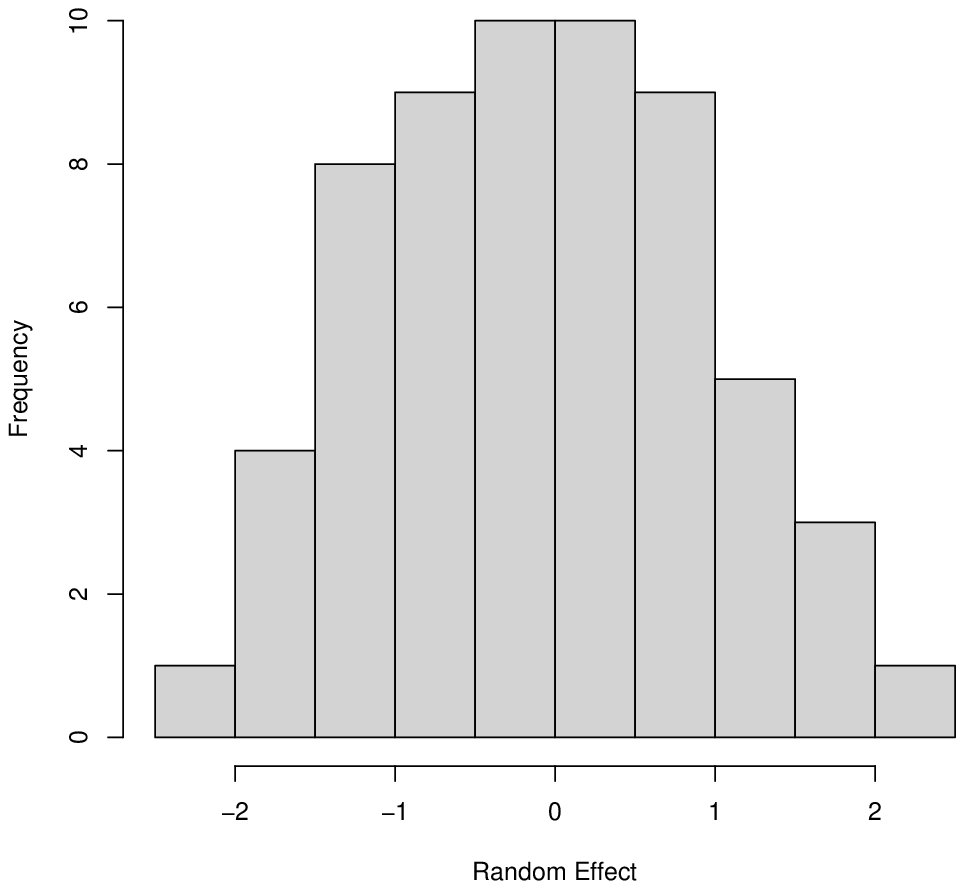}\\
			(a) Spatial Correlation & (b) Random Effects
		\end{tabular}
		
		\caption{Estimated spatial correlation versus distance and histogram of random effects from VI with $\alpha$-divergence, $\alpha = 0.8$ for pine tree data.}
		\label{fig:tree.corr}
	\end{figure}

\begin{figure}
		\centering
		\includegraphics[width=0.98\linewidth]{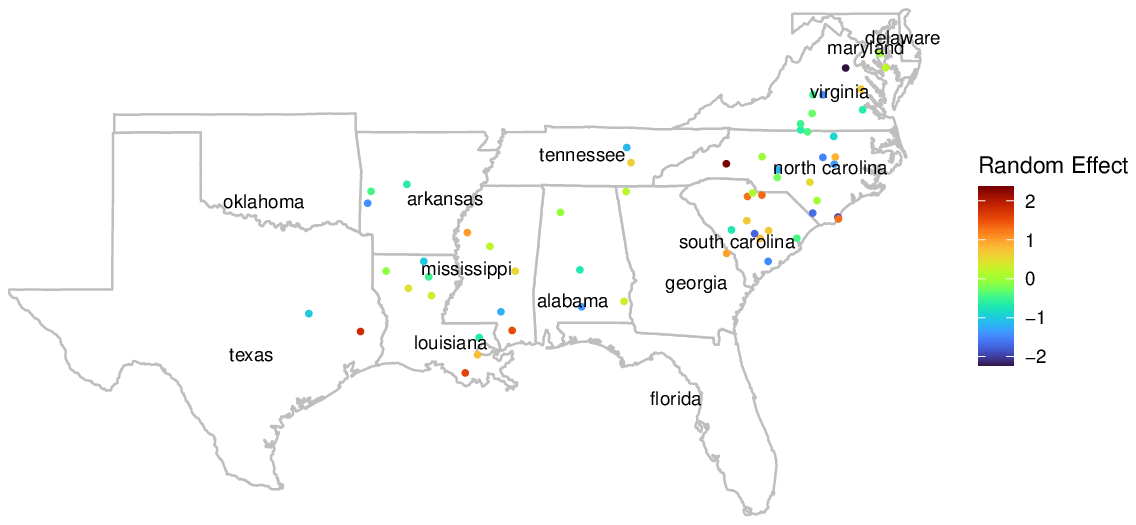}
		\caption{Estimated random effects obtained from VI with $\alpha$-divergence, $\alpha = 0.8$ at the 60 selected locations for the pine tree data. To avoid the overlapping between points, the same jitters are added as in Figure~\ref{fig:tree.mortality}.}
		\label{fig:tree.map}
	\end{figure}

	Since the statistical inference of the $\alpha$-divergence is very close to HMC and it takes less computing time, its inference results are presented in the following.  Figure~\ref{fig:tree.corr}(a) shows the estimated spatial correlation versus the standardized great circle distance. The curve shows that the spatial correlation decays to 0.15 at a distance 0.02. Since the distance matrix is standardized into range $[0,1]$ by dividing the maximum great circle distance between site pairs (1983.41 km), it means that locations within a range of 39.67 km ($0.02 \times1983.41$ km) have correlations greater than or equal to 0.15. Locations that are over 47.6 km have correlations smaller than 0.1, which can be considered as no correlations. Figure~\ref{fig:tree.corr}(b) shows the histogram of estimated random effects, which are in the range of $(-2, 2)$, indicating large spatial variability in survival rate. Figure~\ref{fig:tree.map} shows the estimated spatial random effect at each site on the map. The map shows that moving from Virginia to South Carolina, dots tend to change from light blue/green to orange and brown, indicating the spatial random effect increases and the survival rate decreases. There are also several sites in Texas and Louisiana that have large estimated random effects, indicating locally reduced survival rates.

	\begin{table}
		
		\centering
		\caption{The negative log likelihood and computing time of $\alpha$-divergence, KL divergence, and HMC inferences with pine tree data.}	\label{tbl:tree.summary}
		\begin{center}
			\begin{tabular}{rrrr}
				\hline
				\hline
				& $\alpha = 0.8$ & KL & HMC \\
				\hline
				NLL & 11149.40  & 13712.82& 11147.80 \\
				Time (hours) &3.19& 5.25&  7.79\\
				\hline
				\hline
			\end{tabular}
		\end{center}
	\end{table}
	
	\begin{table}
		
		\caption{The parameter estimates and SE of three inference methods for the pine tree data.}\label{tbl:pars.tree}
		\centering
		\begin{center}
			
			\begin{tabular}{l|rr|rr|rr}
				\hline
				\hline
				\multirow{2}{*}{Parameter}	& \multicolumn{2}{c}{$\alpha$-divergence} &
				\multicolumn{2}{c}{KL divergence} & \multicolumn{2}{c}{HMC}\\
				\cline{2-7}
				& Estimate & SE & Estimate & SE& Estimate & SE  \\
				\hline
				Treatment (Control) & 1.0620 & 0.0458 & 0.7204 & 0.0817 & 1.0648 & 0.0767 \\
				Treatment (Light) &0.2364 & 0.0532 & 0.0728 & 0.0842 & 0.2365 & 0.0800 \\ 
				Physical region (Coastal Plain) & 0.0273 & 0.0501 & $-$0.1729 & 0.1007 & 0.5873 & 0.7354 \\ 
				Physical region (Piedmont) &$-$0.3027 & 0.0556 & $-$0.2214 & 0.1088 & 0.1517 & 0.7535 \\ 
				Crown class (dominant) & $-$2.1722 & 0.0764 & $-$2.5599 & 0.1101 & $-$2.1696 & 0.1163 \\ 
				Crown class (codominant) &$-$2.1563 & 0.0591 &$-$2.6741 & 0.0833 & $-$2.1554 & 0.0840 \\ 
				Crown class (intermediate) &$-$1.1116 & 0.0495 & $-$1.3718 & 0.0599 & $-$1.1052 & 0.0660 \\ 
				DBH & $-$0.2040 & 0.0086 & $-$0.4066 & 0.0223 & $-$0.1929 & 0.0287 \\ 
				TH & $-$0.0211 & 0.0012 & $-$0.0789 & 0.0028 & $-$0.0222 & 0.0040 \\ 
				$a$ & 0.0444 & 0.0023 & 2.1949 & 0.1054 & 0.0243 & 0.0240 \\ 
				$b$ & 0.9157 & 0.0137 & 0.9487 & 0.0484 & 0.9178 & 0.0154 \\ 
				$s_{\gamma}^2$ & 1.4298 & 0.2348 & 4.9903 & 1.0327 & 1.3903 & 0.3621 \\ 
				$\nu$ &  0.0102 & 0.0028 & 0.0425 & 0.1024 & 0.0103 & 0.0029 \\ 
				\hline
				\hline
			\end{tabular}
			
		\end{center}
	\end{table}

	\section{Conclusion and Discussion}\label{sec:vi.con}
	
	In this study, driven by the computational demands for spatial lifetime models with a large number of locations, we focus on the capability and computational efficiency of VI in analyzing two real spatially correlated lifetime datasets. The classic MCMC inference approach often requires extensive efforts on parameter tuning to achieve convergence. The VI method, in contrast, reformulates the inference challenge to an optimization problem. Once the optimized variational parameter is obtained, the model inference can be completed immediately. 
	
	There are several contributions in this work. First, we consider a unified VI framework that comprises a stochastic optimization algorithm and MC approximation for the variational bound. To enhance generality, the R\'enyi’s $\alpha$-divergence, which is the root divergence of the classical KL divergence, is adopted in the framework.  Second, the performance of VI is studied in two spatial lifetime models, the PH model and the CEM, both incorporating time-varying covariates, aiming to demonstrate the versatility of VI across different types of spatial lifetime models. Third, two real datasets, Titan GPU lifetime data and pine tree survival data are used to study the VI inference framework and show the performance of VI with $\alpha$-divergence is comparable to HMC with less computational cost.
	
	Case studies with the Titan GPU lifetime data and pine tree lifetime data show that VI with $\alpha$-divergence has comparable inference performance as HMC.  VI with $\alpha$-divergence has very close NLL values to HMC while speeding up the computing time by around 2 times in both datasets.  The parameter estimation and uncertainty comparisons in two datasets indicate that the KL divergence produces a different posterior from the other two methods. Besides, KL divergence takes a longer time to meet the convergence criteria. A possible explanation is that the parameter space explored by KL divergence is more restricted than $\alpha$-divergence, potentially leading to slower convergence and local optimal solutions. In contrast, $\alpha$-divergence encourages a more flexible variational distribution, resulting in better inference performance. In summary, based on these two applications, we find $\alpha$-divergence with $\alpha<1$ has comparable performance as HMC but with better computational efficiency. 
	
	There are a few aspects to consider when adopting the VI framework for analyzing spatially correlated lifetime data. The first aspect is the choice of the number of samples $h$ in the MC approximation in Equation~\ref{eq:mc.ad.obj}. Choosing the number of samples used in MC is crucial for the approximation quality. Since the proposed VI algorithm relies on the gradient-based optimization, both the variational bound (the objective function) and its gradient need to be well approximated. To select a reasonable $h$, one wants the number of samples $h$ to provide relatively stable estimations of the variational bound and its gradient. While larger $h$ generally improves estimation stability, it also increases computational cost.
	For the GPU and pine tree datasets in this study, we evaluated the trade-off between computational time and approximation stability across different $h$ values.  Based on the variance of the variational bound and its gradient, we selected 
	$h=500$ for the GPU dataset and $h=800$ for the tree dataset. Additional details and results are available in the Supplementary Section 1. In practice, one needs to manually select $h$ by balancing computational efficiency with approximation quality.
	
	The selection of $\alpha$ values in the $\alpha$-divergence is another important aspect to be considered. While cross-validation (CV) would be the ideal approach for selecting $\alpha$, CV is a challenging task with spatially correlated lifetime data. Randomly splitting the dataset into $k$ folds can result in spatial dependencies between the response in the training, testing, and validation sets, potentially leading to overfitting and impacting the choice of $\alpha$-value. In the case studies, we consider a few $\alpha$ values with $\alpha<1$ and observe similar inference performances with both datasets. So, in this paper, $\alpha = 0.8$ is considered. Supplementary Section 2 presents detailed inference results using VI with $\alpha \in [0.3,0.9]$ (with increments of 0.1) for the GPU dataset. The consistent performance across this range suggests that choosing any $\alpha$ value smaller than 1 and not too close to 1 (e.g., $\alpha \leq 0.9$) can yield good inference results. However, developing appropriate CV methods for selecting $\alpha$ in spatially structured datasets remains an interesting direction for future research.
	
	The choice of variational distribution also needs to be considered. In this study, to simplify the optimization in the VI algorithm, we adopt the commonly used mean-field approach, that is, assuming the variational distributions are independent of each other. Several works (e.g., \citeNP{bernardi2024variational}, \citeNP{quiroz2023gaussian}) in the literature also considered independent or conditionally independent variational distributions in spatial-temporal models (e.g., vector autoregressive and state-space models) to balance computational efficiency and estimation accuracy. However, the independence assumption can introduce some estimation bias, especially for the spatial random effects. We evaluate the impact of this choice using the two datasets in this study, and the results are available in Supplementary Section 3. The results show that for data with stronger spatial correlations, independent variational distributions can lead to relatively larger bias in estimating random effects. Nevertheless, residual plots and negative log-likelihood comparisons between $\alpha$-divergence and HMC methods reveal that this assumption does not significantly compromise the $\alpha$-divergence's overall inference capability. While adopting a variational distribution that incorporates spatial dependency structures for the random effects could improve estimation accuracy, it would also introduce computational challenges and additional complexities in the VI estimation process. This represents a trade-off between the accuracy of random effects estimation and the efficiency of optimization. In practice, researchers can choose the variational distribution of the random effects based on the dataset size, model complexity, and the spatial correlation ranges.
	
	There are some future research directions. First, regarding $\alpha$-value selection in $\alpha$-divergence, while our current approach uses empirical testing of different values, developing more systematic methods for $\alpha$ selection would be beneficial. Second, although the main focus of this study is to demonstrate the inference capability of variational inference in spatial survival models, it will be interesting to examine prediction performance for HMC and the VI methods on the testing set. For spatially correlated lifetime data, both research directions require a proper CV design or training-testing set split, where methods reviewed in \citeN{otto2024review} can be applicable.  Works by \citeN{jacquez1999spatial}, \citeN{cressie2003spatial}, and \shortciteN{santi2021handling} also provide insights regarding the desired prior information of spatial dependency when splitting the data into different folds. Additionally, this study focuses on the Bayesian analysis framework. It will be interesting to consider frequentist approaches, like the generalized spatial mixed effect models, as alternatives. Comparing these methods' inference capabilities and computational efficiency with the current VI framework would provide valuable insights. Finally, it may be interesting to consider a mixture variational distribution or other flexible structures to capture the possible dependencies among parameters as well as the multi-mode behavior of posterior distributions.

	\section*{Data Availability Statement}
	The GPU lifetime data in this study are openly available at \url{https://doi.ccs.ornl.gov/dataset/4aa54c30-3d51-5443-b839-88a130e4c713}, and the pine tree lifetime data are openly available at \url{https://www.tandfonline.com/doi/suppl/10.1080/01621459.2014.995793?scroll=top}.
	\bibliographystyle{chicago}
	\bibliography{ref}
	
	
\end{document}